\documentclass[12pt]{amsart}
\usepackage{amsmath}
\usepackage{amsxtra}
\usepackage{amscd}
\usepackage{amsthm}
\usepackage{amsfonts}
\usepackage{amssymb}
\usepackage{eucal}
\def\({\left(}
\def\){\right)}
\newcommand{\bra}[1]{\langle #1 |}        %bra
\newcommand{\ket}[1]{{| #1 \rangle}}      %ket
\newcommand{\nn}{\nonumber}
\newcommand{\bea}{\begin{eqnarray}}
\newcommand{\ena}{\end{eqnarray}}
\def\bel{\begin{eqnarray}}
\def\enl{\end{eqnarray}}
\newcommand{\be}{\begin{eqnarray*}}
\newcommand{\en}{\end{eqnarray*}}
\newcommand{\ba}{\begin{array}}
\newcommand{\ea}{\end{array}}

\def\P{\mathcal P}

\newcommand{\R}{{\mathbb R}}
\newcommand{\C}{{\mathbb C}}
\newcommand{\Z}{{\mathbb Z}}

\newcommand{\cA}{\mathcal{A}}
\newcommand{\cP}{\mathcal{P}}

\newcommand{\slt}{\mathfrak{sl}_2}
\newcommand{\slth}{\widehat{\mathfrak{sl}}_2}

\newcommand{\ad}{{\rm ad}}
\newcommand{\tr}{{\rm tr}}

\newcommand{\Tr}{{\rm Tr}}

\newcommand{\End}{\mathop{\rm End}}

\newenvironment{tenumerate}{
  \begin{enumerate}
  
  }{\end{enumerate}}
\newcommand{\bi}{\begin{tenumerate}}
\newcommand{\ei}{\end{tenumerate}}
\newcommand{\isoto}[1][]%
{{\mathop{\buildrel{\sim}\over\longrightarrow}\limits_{#1}}}

\newcommand{\la}{\lambda}

\newcommand{\e}{\epsilon}

\newcommand{\eb}{\bar{\epsilon}}
\newcommand{\z}{\zeta}

\numberwithin{equation}{section}
\newtheorem{thm}{Theorem}[section]
\newtheorem{prop}[thm]{Proposition}
\newtheorem{lem}[thm]{Lemma}
\newtheorem{rem}[thm]{Remark}

\def\l{\lambda}
\def\s{\sigma}

\def\b#1{{\overline{#1}}}

\begin{document} 
\title[Correlation functions]
{Algebraic representation of correlation functions
in integrable spin chains}
\date{\today}
\author{H.~Boos, M.~Jimbo, T.~Miwa, F.~Smirnov and Y.~Takeyama}
\address{HB: Physics Department, University of Wuppertal, D-42097,
Wuppertal, Germany\footnote{
on leave of absence from 
Skobeltsyn Institute of Nuclear Physics, 
MSU, 119992, Moscow, Russia
}}\email{boos@physik.uni-wuppertal.de}
\address{MJ: Graduate School of Mathematical Sciences, The
University of Tokyo, Tokyo 153-8914, Japan}\email{jimbomic@ms.u-tokyo.ac.jp}
\address{TM: Department of Mathematics, Graduate School of Science,
Kyoto University, Kyoto 606-8502, 
Japan}\email{tetsuji@math.kyoto-u.ac.jp}
\address{FS\footnote{Membre du CNRS}: Laboratoire de Physique Th{\'e}orique et
Hautes Energies, Universit{\'e} Pierre et Marie Curie,
Tour 16 1$^{\rm er}$ {\'e}tage, 4 Place Jussieu
75252 Paris Cedex 05, France}\email{smirnov@lpthe.jussieu.fr}
\address{YT: Graduate School of Pure and Applied Sciences, 
Tsukuba University, Tsukuba, Ibaraki 305-8571, Japan}
\email{takeyama@math.tsukuba.ac.jp}

\dedicatory{To the memory of Daniel Arnaudon}

\begin{abstract}
Taking the XXZ chain as the main example,
we give a review of 
an algebraic representation of correlation functions 
in integrable spin chains obtained recently. 
We rewrite the previous formulas in a form which works
equally well for the physically interesting homogeneous chains. 
We discuss also the case of 
quantum group invariant operators and 
generalization to the XYZ chain. 
\end{abstract}
%%%%%%%%%%%%%%%%%%%%%%%%%%%%%%%%%%%%%
\maketitle
\bigskip

\setcounter{section}{0}
\setcounter{equation}{0}

\section{Introduction}\label{ref:intro}

The investigation of integrable spin chains has a long history
since Bethe's work \cite{Bethe}, in which the Bethe Ansatz method
was invented. 
It was only a start, 
and later was followed by a line 
of new ideas and concepts such as commuting transfer matrices,
the Yang-Baxter equation, the quantum inverse scattering method,
quantum groups and the quantum KZ equation. In a series of papers
\cite{BJMST1,BJMST2,BJMST3,BJMST4}, we studied an algebraic formula
for the correlation functions in the infinite XXX, XXZ and XYZ spin chains.
Our method is a synthesis of those mentioned above.

The study of correlation functions has been a highlight
in the researches of these spin chains. 
In the early days the only knowledge was 
the nearest neighbor correlator, written in terms of $\log2$
for the XXX model.  
It was a big surprise when 
Takahashi \cite{Tak} found $\zeta(3)$ in the next-nearest
correlator, where $\zeta(s)$ is the Riemann zeta function.  
In \cite{JMMN,JM,JM2}, the quantum vertex operators in the representation
theory of the quantum affine $\mathfrak{sl}_2$ algebra were used to obtain
multiple integral formulas for the general correlation functions of the
XXZ model. 
Kitanine, Maillet, Slavnov and Terras 
rederived and further generalized these integral formulas 
to include magnetic field and time \cite{KMT,KMST}
(see \cite{KMST1} for a review). 
Study of the finite temperature case has 
also been launched recently by G{\"o}hmann, Kl{\"u}mper and Seel 
\cite{GKS},   
and progress has been made 
in the calculation of long distance asymptotics of some 
correlators by the Lyon group \cite{KMST2} 
and Korepin, Lukyanov, Nishiyama and Shiroishi \cite{KLNS}. 
However it was not immediately understood why $\zeta(3)$ appears
in the next-nearest correlators.

In \cite{BK1,BK2} Boos and Korepin explicitly performed 
the multiple integrals for the next-nearest case and beyond, 
and again found odd integer values of $\zeta(s)$.
Further exact results including the XXZ chain 
have been obtained by 
Kato, Nishiyama, Sakai, Sato, Shiroishi and Takahashi 
\cite{KSTS,SSNT,KSTS2,SST}. 
In \cite{BKS}, Boos, Korepin and Smirnov studied the inhomogeneous
correlation functions for the XXX model,
and arrived at a conjecture on the algebraic structure
for the general correlation functions: 
in brief, one transcendental function is enough to describe all
of them.  
In the limit of the homogeneous chain, 
the Taylor series expansion of this function produces 
the special values of $\zeta(s)$ as well as $\log 2$. 

In \cite{BJMST1,BJMST2,BJMST3}, we proved the conjecture
by giving an algebraic formula, and obtained 
similar results
in the XXZ and XYZ models. The number of transcendental functions increases to
two and three, respectively, as the number of parameters in the models
increases. The main idea in the proof was the use of the reduced quantum
KZ equation, and the main ingredient in the algebraic formula was
the transfer matrix defined via an auxiliary space of non-integer dimensions.

The algebraic formulas presented in these papers had some
deficiencies: the beauty of the formula was marred by a chip on the edge
of a comb.  
The relevant transfer matrices are `incomplete',  
in that they act on the tensor product where two spaces are omitted. 
Also the formula for the inhomogeneous 
model consists of a sum of terms,  
which have poles when one tries to take the homogeneous limit. 
They cancel each other only after the summation. 
In \cite{BJMST4}, these spots 
were cleaned up in the XXX model.

In the present paper, we give the algebraic formula for the density matrix
in a transparent form, 
not only in the XXX model but also in the XXZ
and the XYZ models. We use the infinite XXZ chain as the main object:
\begin{eqnarray}
H_{XXZ}=\frac12\sum_j(\sigma^1_j\sigma^1_{j+1}+\sigma^2_j\sigma^2_{j+1}+
\Delta\sigma^3_j\sigma^3_{j+1}). 
\label{HAMILTONIAN}
\end{eqnarray}
The density matrix $\rho_n$ belongs to the space
${\rm End}((\C^2)^{\otimes n})^*$ dual to the space of
local operators ${\rm End}((\C^2)^{\otimes n})$.
We consider the space $(\C^2)^{\otimes n}$ as the subchain 
of the entire infinite chain on which the XXZ Hamiltonian acts.
It has the defining property
\begin{eqnarray*}
\rho_n(\mathcal O)=\langle{\rm vac}|\mathcal O|{\rm va}c\rangle.
\end{eqnarray*}
Here the right hand side is the ground state average of the operator
$\mathcal O\in{\rm End}((\C^2)^{\otimes n})$. We will give the formula for
$\rho_n$ in the form
\begin{eqnarray}
\rho_n(\mathcal O)=\frac1{2^n}{\rm tr}_{(\C^2)^{\otimes n}}
(e^{\Omega_n^*}\mathcal O),
\label{OMEGA}
\end{eqnarray}
where $\Omega_n^*$ is a nilpotent 
linear operator acting on 
${\rm End}\bigl((\C^2)^{\otimes n}\bigr)$.
The formula for $\Omega_n^*$ is given by a twofold integral:
\begin{eqnarray}
&&\Omega_n^*
=\frac1{2\kappa^2}
\int\!\!\int\frac{d\mu_1}{2\pi i}\frac{d\mu_2}{2\pi i}
{\rm tr}_{\C^2\otimes\C^2}\left(
B(\mu_{1,2})(1\otimes\pi^{(1)}({\mathcal T}_n^*(\mu_2)))
(\pi^{(1)}({\mathcal T}_n^*(\mu_1))\otimes1)\right)
\label{INT}\\
&&
\times(\omega_1(\mu_{1,2}){\mathcal X}_{1,n}^*(\mu_1,\mu_2)+
\omega_2(\mu_{1,2}){\mathcal X}_{2,n}^*(\mu_1,\mu_2)).\nonumber
\end{eqnarray}
Here, $\kappa$ is a constant, $B(\mu_{1,2})$ is a $4\times4$ matrix
depending on $\mu_{1,2}=\mu_1-\mu_2$, 
and $\omega_i$ ($i=1,2$) are certain transcendental functions.
The operator ${\mathcal T}_n^*(\mu)$ is given 
in terms of the $L$ operator 
$L(\mu)\in U_q(\mathfrak{sl}_2)\otimes{\rm End}(\C^2)$  
as the monodromy matrix in the adjoint action:
\begin{eqnarray*}
{\mathcal T}_n^*(\mu)(\mathcal O)=
L_1(\mu)^{-1}\cdots L_n(\mu)^{-1}\mathcal O L_n(\mu)\cdots L_1(\mu)
\in U_q(\mathfrak{sl}_2)\otimes{\rm End}((\C^2)^{\otimes n}).
\end{eqnarray*}
The deformation parameter $q=e^{\pi i\nu}$
and the anisotropy parameter $\Delta$ in (\ref{HAMILTONIAN}) are related as
$\Delta=\frac{q+q^{-1}}2.$
We denote the irreducible two-dimensional representation of
$U_q(\mathfrak{sl}_2)$ by $\pi^{(1)}$.

The operators ${\mathcal X}_{i,n}^*(\mu_1,\mu_2)$ ($i=1,2$) is obtained
from the monodromy matrix
\begin{eqnarray*}
{\rm Tr}_{\mu_{1,2}}{\mathcal T}_n^*\left(\frac{\mu_1+\mu_2}2\right)
={\mathcal X}_{1,n}^*(\mu_1,\mu_2)-\mu_{1,2}{\mathcal X}_{2,n}^*(\mu_1,\mu_2),
\end{eqnarray*}
where ${\rm Tr}_d$ denotes the $d$-dimensional trace.

The above formula is in the homogeneous case, and the integrand has
poles at $\mu_i=0$ ($i=1,2$). The integral means taking residues
at these poles. A similar formula is also given in the inhomogeneous case
where the Hamiltonian is replaced with the transfer matrix for
the inhomogeneous six vertex model with the spectral parameters
$\lambda_1,\ldots,\lambda_n$ associated with the tensor components 
of 
$(\C^2)^{\otimes n}$. 
For the details, see Theorem \ref{prop:main}
and \eqref{eq:new2}--\eqref{eq:new3}.
In this case, the integrand has poles at
$\mu_i=\lambda_j$ ($i=1,2;j=1,\ldots,n$). 
Taking residues at these poles
we get the formula obtained in the previous paper \cite{BJMST2}.

For a general local operator $\mathcal O$, 
we need two functions $\omega_i$ ($i=1,2$)
to express its expected value. 
However, if $\mathcal O$
is invariant under the action of
$U_q(\mathfrak{sl}_2)$, the formula simplifies, and we need only $\omega_1$.
This case is related to the spin chain with an 
open boundary condition
given by the Pasquier-Saleur Hamiltonian \cite{PS}. The XXZ Hamiltonian
with periodic boundary condition corresponds to the CFT with the central charge
$c=1$. In contrast, the Pasquier-Saleur Hamiltonian corresponds to the CFT
with $c=1-6\nu^2/(1-\nu)$. The above property of the invariant operators
was conjectured in \cite{BKS2}. We give a proof to this conjecture.

We also give a formula similar 
to (\ref{OMEGA}), (\ref{INT}) for the XYZ model.

The paper is organized as follows. In Section \ref{sec:density},
the density matrix is defined. In Section \ref{sec:XXZansatz}, 
an algebraic formula
of the operator $\Omega_n$, which is dual to $\Omega^*_n$, 
is given.
In Section \ref{sec:alternative}, 
the algebraic formula is written
in an alternative form. In Section \ref{sec:invariant}, 
the formula for the invariant operators
are given. 
In Section \ref{sec:XYZ}, 
the formula for the XYZ model is given. 

The text is followed by three appendices. 
In Appendix \ref{sec:review}, we give the 
derivation of the new formula for $\Omega_n$. 
In Appendix \ref{app:book}, we make a
comparison between different conventions 
used in this paper and in the book \cite{JM}. 
In Appendix \ref{app:rho}, 
formulas for the normalization factors are gathered
for the XXZ and the XYZ models.

\section{Density matrix for the XXZ chain}\label{sec:density}

Consider the XXZ Hamiltonian 
\begin{eqnarray}
H_{\rm XXZ}=\frac{1}{2}\sum_{k=-\infty}^{\infty}
\left( 
\sigma_{k}^1\sigma_{k+1}^1+
\sigma_{k}^2\sigma_{k+1}^2+
\Delta\sigma_{k}^3\sigma_{k+1}^3
\right), 
\label{eq:XXZ}
\end{eqnarray}
where $\sigma^{\alpha} \, (\alpha=1,2,3)$ are 
the Pauli matrices and
\be
\Delta=\cos\pi\nu
\en
is a real parameter. 
We consider the two regimes, the massive regime $\Delta>1$, $\nu\in i \R_{>0}$, 
and the massless regime $|\Delta|<1$, $0<\nu<1$.  

Take a sub-interval of the lattice consisting of sites $1,\ldots,n$,  
where $n$ is a positive integer. 
Let $(E_{\e,\bar{\e}})_j$ denote the matrix unit 
$\left(\delta_{a\epsilon}\delta_{b\bar{\epsilon}}\right)_{a,b=\pm}$
acting on the site $j$. 
By a density matrix, we mean the one 
whose entries are the ground state averages of products
of the $(E_{\e,\bar{\e}})_j$'s, 
\bel
&&\rho_n(\l_1,\ldots,\l_n)\label{DENSITY}\\
&&=\sum_{\e_,\ldots,\e_n\atop{\b\e}_{1},\ldots,{\b\e}_{n}}
{}_{\l_1,\ldots,\l_n}\langle\hbox{vac}|
(E_{{\b\e}_{1},\e_1})_1\cdots(E_{{\b\e}_{n},\e_n})_n|\hbox{vac}\rangle
_{\l_1,\ldots,\l_n}
(E_{\e_1,{\b\e}_{1}})_1\cdots(E_{\e_n,{\b\e}_{n}})_n. 
\nonumber
\enl
Here we consider the model with inhomogeneities 
$\la_1,\cdots,\la_n$ attached to each site. 
More precisely, we mean the following. 

Let 
$V=\C^2$ be the two dimensional vector space 
with basis $v_+,v_-$. 
Throughout this paper, we set 
\begin{equation}
q=e^{\pi i\nu}.
\label{eq:qnu}
\end{equation}
Denote the standard trigonometric $R$ matrix by 
\bea
&&R(\l)=\frac{\rho(\la)}{[\la+1]}\,r(\la),
\label{eq:def-R}\\
&&r(\l)=\begin{pmatrix}
[\l+1]&0&0&0\\
0&[\l]&1&0\\
0&1&[\l]&0\\
0&0&0&[\l+1]
\end{pmatrix}\quad \in\End(V\otimes V). 
\nonumber
\ena
Here the entries are arranged in the order $(++),(+-),(-+),(--)$, and 
\be
[\l]=\frac{q^\l-q^{-\l}}{q-q^{-1}}.
\en
The factor $\rho(\la)=\rho(\la,2)$ will be given later 
(see \eqref{eq:L} and \eqref{eq:rho_trig1}, \eqref{eq:rho_trig2}). 
Introduce an auxiliary space $V_a\simeq V$ with spectral parameter $\l$,   
and denote by $R_{a,j}$ the $R$ matrix acting on $V_a\otimes V_j$.
Using \eqref{eq:def-R}, we consider the transfer matrix of the inhomogeneous 
six vertex model 
\bea
&&\tr_{V_a}\{R_{a,L}(\l-\l_L)\cdots
R_{a,n}(\l-\l_n)\cdots R_{a,1}(\l-\l_1)\cdots
R_{a,-L}(\l-\l_{-L})\}
\label{eq:6v}
\ena
which acts on the tensor product
\be
V_{-L}\otimes\cdots\otimes V_1
\otimes\cdots\otimes V_n
\otimes\cdots\otimes V_L. 
\en
With each $V_j$ we associate a spectral parameter $\la_j$, 
assuming for definiteness that $\la_j=0$ for $j\le 0$ or 
$j\ge n+1$. 
Let 
$|{\rm vac}\rangle^{(L)}_{\scriptscriptstyle\l_1,\ldots,\l_n}$ 
denote the eigenvector of \eqref{eq:6v} corresponding to the lowest eigenvalue. 
We denote the dual eigenvector by
${}_{\scriptscriptstyle\l_1,\ldots,\l_n}\hskip-12pt{}^{(L)}\langle{\rm vac}|$,
normalized so that 
${}_{\scriptscriptstyle\l_1,\ldots,\l_n}\hskip-12pt{}^{(L)}\langle{\rm vac}
|{\rm vac}\rangle^{(L)}_{\scriptscriptstyle\l_1,\ldots,\l_n}=1$.
The vacuum expectation value in \eqref{DENSITY} is defined to be the thermodynamic limit 
\be
&&{}_{\scriptscriptstyle\l_1,\ldots,\l_n}
\langle\hbox{vac}|(E_{{\b\e}_{1},\e_1})_1\cdots
(E_{{\b\e}_{n},\e_n})_n|\hbox{vac}\rangle
_{\scriptscriptstyle\l_1,\ldots,\l_n}\\
&&=\lim_{L\rightarrow\infty}
{}_{\scriptscriptstyle\l_1,\ldots,\l_n}\hskip-12pt{}^{(L)}\langle{\rm vac}|
(E_{{\b\e}_{1},\e_1})_1\cdots(E_{{\b\e}_{n},\e_n})_n
|{\rm vac}\rangle^{(L)}_{\scriptscriptstyle\l_1,\ldots,\l_n}.
\en

For an arbitrary local operator $\mathcal O\in\End(V^{\otimes n})$, we have 
\bel
{}_{\l_1,\ldots,\l_n}
\langle{\rm vac}|\mathcal O|{\rm vac}\rangle_{\l_1,\ldots,\l_n}
=\tr_{V^{\otimes n}}
({\mathcal O}\rho_n). 
\label{ANOTHER}
\enl
Our aim is to give an algebraic representation for the
density matrix $\rho_n$. 

\section{Algebraic formula}\label{sec:XXZansatz}

The density matrix $\rho_n$ is an operator on $V^{\otimes n}$. 
To present the result, let us pass from operators to 
vectors in $V^{\otimes 2n}$. 
We number the spaces as 
\bea
V_1\otimes\cdots\otimes V_n\otimes V_{\b n}
\otimes\cdots\otimes V_{\b1}.   
\label{eq:vvbar}
\ena
We use the following convention for the indices: 
for example, if $u=\sum u'\otimes u'',v=\sum v'\otimes v''$
are vectors in $V\otimes V$, then we write 
\be
u_{1,\b1}v_{\b2,2}=
\sum u'\otimes v''\otimes v'\otimes u''
\quad \in
V_1\otimes V_2\otimes V_{\b2}\otimes V_{\b1}.
\en
Similarly, we indicate by suffix the tensor components on which 
operators act non-trivially. 

Introduce a function $h_n$ with values in \eqref{eq:vvbar}
\footnote{There is an erratum in \cite{BJMST2}; 
the right hand side of the formula seven lines below (13.1) should read 
$\prod_{j=1}^n
(-{\b\e}_{j})\langle\hbox{vac}|(E_{-{\b\e}_{1},\e_1})_1\cdots
(E_{-{\b\e}_{n},\e_n})_n|\hbox{vac}\rangle$. }:
\bea
&&h_n(\l_1,\ldots,\l_n)
=\sum
\prod_{j=1}^n
(-{\b\e}_{j})\langle\hbox{vac}|(E_{-{\b\e}_{1},\e_1})_1\cdots
(E_{-{\b\e}_{n},\e_n})_n|\hbox{vac}\rangle
\label{eq:defh}\\
&&\hskip20pt\times v_{\e_1}\otimes\cdots\otimes v_{\e_n}\otimes
v_{{\b\e}_{ n}}\otimes\cdots\otimes v_{{\b\e}_{1}}.
\nonumber
\ena
In Section \ref{sec:alternative}, 
we discuss more about the transition from $\rho_n$ to $h_n$, 
and vice versa. 
We mention here only that spectral parameters $\la_j$, $\la_j+1$
are attached to the spaces $V_j$ and $V_{\bar j}$, respectively. 

The function $h_n$ is known to 
satisfy the following system of equations: 
\begin{eqnarray}
&& 
h_{n}(\ldots , \lambda_{k+1}, \lambda_{k}, \ldots )=
\check{R}_{k, k+1}(\lambda_{k, k+1})
\check{R}_{\overline{k+1}, \overline{k}}(\lambda_{k+1, k})
h_{n}(\ldots , \lambda_{k}, \lambda_{k+1}, \ldots )\,, 
\label{eq:rqkz1} \\ 
&& 
h_{n}(\lambda_{1}-1, \lambda_{2}, \ldots , \lambda_{n})=
A_{n}(\lambda_{1}, \ldots , \lambda_{n})
h_{n}(\lambda_{1}, \lambda_{2}, \ldots , \lambda_{n})\,, 
\label{eq:rqkz2}\\
&& 
\mathcal{P}_{1, \b1}^{-}h_{n}(\lambda_{1}, \ldots , \lambda_{n})
=\frac12s_{1, \b1}
h_{n-1}(\lambda_{2}, \ldots , \lambda_{n})_
{2, \ldots , n, \b n, \ldots , \b2}\,,
\label{eq:rqkz3}\\ 
&& 
\mathcal{P}_{n, \b n}^{-}h_{n}(\lambda_{1}, \ldots , \lambda_{n})
=\frac12s_{n, \b n}
h_{n-1}(\lambda_1, \ldots , \lambda_{n-1})_
{1, \ldots , n-1, \b{n-1}, \ldots , \b1}\,.
\nonumber
\end{eqnarray}
Here the notation is as follows. We set 
$\lambda_{i,j}=\lambda_{i}-\lambda_{j}$, 
$\check R=P\, R$ with $P$ being the transposition, 
\be
\cP^-=\frac{1}{2}(I-P)
\en
is the projection onto $\C s$ where $s$ denotes the vector 
\bea
&&
s=v_+\otimes v_--v_-\otimes v_+\in V\otimes V, 
\label{eq:sing}
\ena
and 
\be
A_{n}(\lambda_{1}, \ldots , \lambda_{n})=
(-1)^{n} P_{1,\bar{1}}
R_{1, \bar{2}}(\lambda_{1,2}-1) \cdots 
R_{1, \bar{n}}(\lambda_{1,n}-1)
R_{1,n}(\lambda_{1, n}) \cdots R_{1, 2}(\lambda_{1,2}). 
\en

We call \eqref{eq:rqkz1}--\eqref{eq:rqkz3} reduced qKZ (rqKZ) equations. 
We are going to construct a solution of these equations in a 
certain specific form. 
For that purpose, we will need three ingredients: 
monodromy matrix, trace functional ${\rm Tr}_\la$, 
and transcendental functions $\omega_1(\la,\nu),\omega_2(\la,\nu)$. 
Let us explain them. 

Let $E,F,H$ be the standard generators of $U_q(\slt)$. 
We consider the $L$ operator with auxiliary space of dimension $d$, 
\bea
&&L(\l)=\frac{\rho(\l,d)}{[\l+\frac d2]}\ell(\l),
\label{eq:L}
\ena
where
\be
\ell(\l)=
\begin{pmatrix}
[\l+\frac{1+H}2]&Fq^{\frac{H-1}2}\\
q^{\frac{1-H}2}E&[\l+\frac{1-H}2]
\end{pmatrix}
\in U_q(\slt)\otimes\End(V).
\en
The normalization factor $\rho(\l,d)$ is chosen to satisfy 
\be
L(\l)L(-\l)&=&1\otimes I_V\quad\hbox{(unitarity relation)},\\
\s^2L(\l)^t\s^2&=&-L(-1-\l)\quad\hbox{(crossing symmetry)}.
\en
Here $L(\l)^t$ is the transposed matrix with respect to $\End(V)$. 
For the explicit formula of $\rho(\la,d)$,  
see \eqref{eq:rho_trig1}, \eqref{eq:rho_trig2}. 
We have, in particular,
\bel
\frac{\rho(\l,d)}{[\l+\frac d2]}\frac{\rho(\l-1,d)}{[\l+\frac d2-1]}=
-\frac1{[\l-\frac d2][\l+\frac d2]}.
\label{POLE}
\enl
We define the monodromy matrix $T_n(\la)=T_n(\la|\l_1,\ldots,\l_n)$ by
\bea
&&T_n(\la)=L_{\bar 1}(\la-\la_1-1)\cdots L_{\bar n}(\la-\la_n-1)
L_n(\la-\la_n)\cdots L_1(\la-\la_1).
\label{eq:T}
\ena

The trace functional ${\rm Tr}_\l$ is the composition map 
\bea
&&{\rm Tr}_\l~:~U_q(\slt)\rightarrow
U_q(\slt)/[U_q(\slt),U_q(\slt)]\rightarrow
\lambda\C[\zeta,\zeta^{-1}]\oplus\C[\zeta,\zeta^{-1}],  
\label{eq:trfcn}
\ena
where $\zeta=q^\lambda$.  
The first map is the canonical map, 
and the second is defined by setting for any $m\in\Z$ 
\be
&&{\rm Tr}_\l(q^{mH})=
\begin{cases}
[m\l]/[m]&\hbox{ if }m\not=0;\\
\l&\hbox{ if }m=0,
\end{cases}
\en
and for any $x \in U_{q}(\slt)$ 
\begin{eqnarray*}
{\rm Tr}_{\lambda}(Cx)=\frac{q^{\lambda}+q^{-\lambda}}{(q-q^{-1})^{2}}\, 
{\rm Tr}_{\lambda}(x), 
\end{eqnarray*}
where $C$ is a central element given by 
\begin{eqnarray*}
C=\frac{q^{-1+H}+q^{1-H}}{(q-q^{-1})^{2}}+EF. 
\end{eqnarray*}
An equivalent way of defining ${\rm Tr}_\l\,x$ 
for $x\in U_q(\slt)$ is as follows.   
It is the unique element of 
$\lambda\C[\zeta,\zeta^{-1}]\oplus\C[\zeta,\zeta^{-1}]$ 
such that, for all $k+1$ dimensional   
irreducible representation 
$\pi^{(k)}~:~U_q(\slt)\to \End(\C^{k+1})$  
we have 
\be
\left({\rm Tr}_\l\,x\right)|_{\l=k+1}=\tr_{V^{(k)}}\pi^{(k)}(x)
\qquad (k\in\Z_{\geq0}).  
\en

With this definition of ${\rm Tr}_\la$, the 
`trace' of the monodromy matrix has a unique decomposition 
\bea
&&{\rm Tr}_{\mu_{12}}T_n\left(\frac{\mu_1+\mu_2}{2}\right)
\label{eq:GGt}\\
&&
=X_{1,n}(\mu_1,\mu_2|\l_1,\ldots,\l_n)
-\mu_{1,2}X_{2,n}(\mu_1,\mu_2|\l_1,\ldots,\l_n),
\nonumber
\ena
where 
$X_{i,n}(\mu_1,\mu_2)=X_{i,n}(\mu_1,\mu_2|\la_1,\cdots,\la_n)$
($i=1,2$) are matrices whose entries are 
rational functions in the variables
$q^{\mu_1},q^{\mu_2},q^{\la_1},\cdots,q^{\la_n}$. 
Note that, with the substitution $\l=\frac{\mu_1+\mu_2}2-\l_j,d=\mu_{1,2}$, 
the right hand side of (\ref{POLE}) becomes
\bel
-\frac1{[\mu_1-\l_j][\mu_2-\l_j]}\,.  
\label{POLE2}
\enl

Finally, define the functions $\omega_i(\l,\nu)$ ($i=1,2$)
\footnote{Our $\omega_1$, $\omega_2$ here are denoted 
$\omega$ and $\tilde{\omega}$ in \cite{BJMST2}.} 
by
\begin{equation}
\kappa\,d\log\varphi(\l,\nu)=\omega_1(\l,\nu)d(\l\nu)+\omega_2(\l,\nu)d\nu, 
\label{eq:omegaXXZ}
\end{equation}
where
\be
\varphi(\l,\nu)=\rho(\l)
\left(\frac{[\l-1]}{[\l+1]}\right)^{1/4},
\quad 
\kappa=\frac{\sin\pi\nu}{\pi\nu},
\en
and $\nu$ is given in \eqref{eq:qnu}. 
The function $\rho(\la)=\rho(\la,2)$ depends also on $\nu$ 
and is defined in \eqref{eq:rho_trig1}, \eqref{eq:rho_trig2} 
in each regime. 

We are now in a position to state the algebraic formula. 
Set 
\bea
&&{\bf s}_n=\prod_{j=1}^ns_{j,\b j}.
\label{eq:sn}
\ena

\begin{thm}\label{prop:main}
The following formula gives a solution of the 
rqKZ equations \eqref{eq:rqkz1}--\eqref{eq:rqkz3}: 
\bel
&&h_n(\la_1,\cdots,\la_n)
=\frac1{2^n}e^{\Omega_n(\la_1,\cdots,\la_n)}\,{\bf s}_n,
\quad
\label{eq:ansatz}
\enl
where
\bel
&&\quad\Omega_n(\la_1,\cdots,\la_n)
=\frac{(-1)^n}{2\kappa^2}\int\!\!\int
\frac{d\mu_1}{2\pi i}\frac{d\mu_2}{2\pi i}
\label{eq:newformula}\\
&&\times
\left(\omega_1(\mu_{1,2})X_{1,n}(\mu_1,\mu_2)
+\omega_2(\mu_{1,2})
X_{2,n}(\mu_1,\mu_2)\right)
{\rm Tr}_{2,2}\left(T_n(\mu_1)\otimes T_n(\mu_2)\cdot
B(\mu_{1,2})\right)
\nn
\enl
and
\bea
B(\mu)=\frac{1}{2}\frac{[\mu]}{[\mu-1][\mu+1]}
\begin{pmatrix}
0 &  &  &  \\
  & q^\mu+q^{-\mu}&-q-q^{-1}&  \\
  &-q-q^{-1}& q^\mu+q^{-\mu}&\\
  &  &      &0 \\
\end{pmatrix}.
\label{eq:Aab}
\ena
Here $\omega_i(\la,\nu)$ are given in \eqref{eq:omegaXXZ}, 
$X_{a,n}$ are defined by \eqref{eq:GGt},
${\rm Tr}_{2,2}(x\otimes y)= {\rm Tr}_{2}(x){\rm Tr}_2(y)$,
and the matrix $B(\mu)$ acts on the auxiliary space $\C^2\otimes\C^2$. 
The integral in $\mu_i$ ($i=1,2$) means taking residues at
the poles $\mu_i=\l_1,\cdots,\l_n$ 
which result from (\ref{POLE2}).

In the massive regime, $h_n$ coincides with the vector form of 
the density matrix. 
\end{thm}
Conjecturally the same formula 
gives also   
the density matrix in the massless regime as well. 

In the above formula, we abbreviated all the variables 
$\nu,\l_1$, etc., other than $\mu_1,\mu_2$.  
This formula was given earlier in \cite{BJMST2} in a different form. 
Since the integrand has no pole at $\la_i=\la_j$,  
the present formula 
is equally valid in the homogeneous chain where $\l_1=\cdots=\l_n=0$.
We show the equivalence of the two formulas 
in Appendix \ref{sec:review}.  

There is another representation for the operator $\Omega_n$   
using the `invariant' trace 
\bea
{\rm Tr}_\lambda^q(A)={\rm Tr}_\lambda (q^{-H}A).  
\label{eq:invtr}
\ena
Define $X^q_{1,n},X^q_{2,n}$ and $B^q(\mu)$ by  
\bea
&&{\rm Tr}_{\mu_{12}}^q
T_n\left(\frac{\mu_1+\mu_2}{2}\right)
\label{eq:GGt2}\\
&&
=X^q_{1,n}(\mu_1,\mu_2|\l_1,\ldots,\l_n)
-\mu_{1,2}X^q_{2,n}(\mu_1,\mu_2|\l_1,\ldots,\l_n),
\nonumber\\
&&
B^q(\mu)=\frac{[\mu]}{[\mu-1][\mu+1]}
\begin{pmatrix}
0 &  &  &  \\
  & q&-q^{-\mu}&  \\
  &-q^{\mu}&q^{-1}& \\
  &  &      &0 \\
\end{pmatrix}.
\label{eq:Bsigma}
\ena
Then $\Omega_n$ can also be written as \eqref{eq:newformula}, 
with $X_{a,n}(\mu_1,\mu_2)$ and $B(\mu)$ replaced by 
$X^q_{a,n}(\mu_1,\mu_2)$ and $B^q(\mu)$, respectively. 

The existence of this second representation 
is a peculiar feature of the XXZ model which has analogs 
neither in the XXX nor in the XYZ models. 

\begin{rem}
The choice of the operator $B(\mu)$ in \eqref{eq:Aab} is not unique. 
For example, there is a freedom of adding identity to $B(\mu)$ 
(see Lemma \ref{lem:app-main2}). 
\end{rem}

\section{An alternative representation}\label{sec:alternative}

In this section we return from vectors in $V^{\otimes 2n}$ 
to operators on $V^{\otimes n}$. 
Let us recall some generalities concerning 
the action of quantum groups on these spaces. 

Denote by $\pi_\l:U_q'(\slth)\rightarrow U_q(\slt)$ 
the evaluation homomorphism 
\be
&&\pi_\l(e_0)=q^\l F,\quad
\pi_\l(f_0)=q^{-\l} E,\quad
\pi_\l(q^{\pm h_0/2})=q^{\mp H/2},\\
&&\pi_\l(e_1)=q^\l E,\quad
\pi_\l(f_1)=q^{-\l} F,\quad
\pi_\l(q^{\pm h_1/2})=q^{\pm H/2}.
\en
For the representation $\pi^{(k)}$ of $U_q(\slt)$, 
we set $\pi^{(k)}_\l=\pi^{(k)}\circ\pi_\l$. 
We use the coproduct 
\be
\Delta(e_i)&=&e_i\otimes1+q^{h_i}\otimes e_i,
\\
\Delta(f_i)&=&f_i\otimes q^{-h_i}+1\otimes f_i 
\en
to define the action of $U_q'(\slth)$ on a 
tensor product of representations.  

Quite generally, for a finite dimensional $U'_q(\slth)$ module $W$, 
its dual vector space $W^*$ has two module structures defined 
via the antipode $S$ as 
\be
\langle x u,v\rangle =\langle u, S^{\pm 1}(x) v\rangle
\qquad (x\in U_q(\slth), u\in W^*, v\in W). 
\en
Denote these structures by $W^{* S^{\pm 1}}$. 
We have canonical isomorphisms 
\be
(W^{* \phi})^{* \phi^{-1}}\simeq W,
\quad 
(W_1\otimes W_2)^{* \phi}\simeq W_2^{*\phi}\otimes W_1^{*\phi}  
\en
for $\phi=S^{\pm 1}$. 
The canonical pairing $W^{* S}\otimes W\to \C$ is $U'_q(\slth)$-linear. 
We regard $\End(W)$ as a $U'_q(\slth)$-module via 
\be
\End(W)\simeq W\otimes W^{* S}.
\en
Using the trace ${\rm tr}_W(AB)$, $\End(W)$
may be identified with its dual space. 
The induced dual module structure becomes 
\be
\End(W)^{*S^{-1}}\simeq (W\otimes W^{* S})^{*S^{-1}}\simeq W\otimes W^{* S^{-1}}.
\en

We are mainly concerned with the $2$ dimensional module $V$ 
where the generators $E,F,H$ act 
in the basis $v_+,v_-$ as
\be
E=\begin{pmatrix}0&1\\0&0\end{pmatrix},\quad
F=\begin{pmatrix}0&0\\1&0\end{pmatrix},\quad
H=\begin{pmatrix}1&0\\0&-1\end{pmatrix}. 
\en
Use the letter $V(\la)$ to indicate the evaluation module structure 
$\pi^{(1)}_\l$ on $V$. 
We have then an isomorphism of $U'_q(\slth)$-modules 
\be
V(\la)^{* S^{\pm 1}}\simeq
V(\la\mp 1),
\quad v^*_\epsilon~\mapsto~\epsilon\,v_{-\epsilon},  
\en
where 
$\langle v^{*}_\epsilon, v_{\epsilon'}\rangle=\delta_{\epsilon,\epsilon'}$. 
In particular, the identity operator $I_V\in \End(V)$ 
corresponds to $s\in V\otimes V$ given in \eqref{eq:sing}. 

We started from the tensor product 
\be
\mathcal{S}_n=V_1\otimes\cdots\otimes V_n, 
\qquad V_j=V(\la_j), 
\en
corresponding to the finite interval $1,\ldots,n$ on the lattice.  
Our space of local operators is the $U'_q(\slth)$-module 
\bea
\mathcal{L}_n&=&\End(\mathcal{S}_n)
\label{eq:local}\\
&\simeq &
V(\la_1)\otimes\cdots\otimes V(\la_n)\otimes 
V(\la_{n}-1)\otimes\cdots \otimes V(\la_1-1) 
\nonumber
\ena
on which $x\in U'_q(\slth)$ operates by the 
adjoint action 
\be
\ad\,x\bigl(\mathcal{O}\bigr)=\sum x_i'\mathcal{O}S(x_i'')
\qquad (\mathcal{O}\in\mathcal{L}_n),  
\en
where $\Delta(x)=\sum x_i'\otimes x_i''$. 
In contrast, density matrix belongs to the dual module
\be
\mathcal{L}^*_n&=&\End(\mathcal{S}_n)^{*S^{-1}}
\\
&\simeq &
V(\la_1)\otimes\cdots\otimes V(\la_n)\otimes 
V(\la_{n}+1)\otimes\cdots \otimes V(\la_1+1). 
\en
The action of $x\in U'_q(\slth)$ is, 
in the same notation as above, 
\be
\ad'x\bigl(\mathcal{O}^*\bigr)=\sum x_i''\mathcal{O}^*S^{-1}(x_i')
\qquad (\mathcal{O}^*\in\mathcal{L}_n^*). 
\en
The vector $h_n$ \eqref{eq:defh}
is nothing but the image of the density matrix $\rho_n$ under the latter 
identification. 

In passing we note that,  
for a Hopf subalgebra $U$ of $U'_q(\slth)$, 
$A\in\End(W)$ belongs to the trivial representation
(i.e., $\ad x\bigl(A\bigr)=\epsilon(x) A$ for all $x\in U$
where $\epsilon$ is the counit) 
if and only if $x\cdot A=A\cdot x$ for all $x\in U$. 
In this case we say $A$ is invariant under $U$. 

With this preparation, let us 
rewrite our main formula in the matrix formulation. 
Suppose $\mathcal O^*\in\mathcal{L}_n^*$ and
$v\in V(\lambda_1)\otimes\cdots V(\lambda_n)\otimes V(\lambda_n+1)\cdots
V(\lambda_1+1)$ are identified. 
Then the action $v\mapsto L_i(\mu)v$ is translated to the left multiplication
\be
\mathcal{O}^*&\mapsto&L_i(\mu)\mathcal{O}^*,
\en
while $v\mapsto L_{\bar i}(\mu-1)v$ is translated to 
\be
\mathcal{O}^*&\mapsto&
-\mathcal{O}^*L_i(\mu)^{-1}.
\en
In view of the cyclicity of the trace, 
the action of the `transfer matrix' \eqref{eq:GGt} 
on $v$ turns into 
$$
(-1)^n\Tr_{\mu_{12}}\mathcal{T}_n(\mu)(\mathcal{O}^*), 
$$
where $\mu=(\mu_1+\mu_2)/2$ and 
\be
\mathcal{T}_n(\mu)(\mathcal{O}^*)=
L_n(\mu-\la_n)\cdots L_1(\mu-\la_1)\cdot 
\mathcal{O}^*\cdot L_1(\mu-\la_1)^{-1}\cdots L_n(\mu-\la_n)^{-1}.  
\en
Notice that in this formula the normalization factor of the 
$L$ operator cancels out.  
Regard the operator $\Omega_n$ as acting on $\mathcal{L}_n^*$ 
via the above formula.  
Then the density matrix can be written as  
$$
\rho _n=\frac{1}{2^n}e^{\Omega_n}(I), 
$$
where $I$ is the identity operator. 

Similarly, denote by $\Omega_n^*$ 
the operator corresponding to $\Omega_n$, acting 
on the space of local operators $\mathcal{L}_n$.  
Then the main formula \eqref{eq:ansatz} can be rewritten as 
\bea
\langle \text{vac}|\mathcal{O}|\text{vac}\rangle=\frac 1 {2^n}
\text{tr}_{V^{\otimes n}}\(e^{\Omega^*_n}(\mathcal{O})\), 
\label{eq:new2}
\ena
where 
\bea
&&\Omega^*_n(\la_1,\cdots,\la_n)
=\frac{1}{2\kappa^2}\int\!\!\int
\frac{d\mu_1}{2\pi i}\frac{d\mu_2}{2\pi i}
{\rm Tr}_{2,2}\left(B(\mu_{1,2})
(I\otimes \mathcal{T}^*_n(\mu_2))
(\mathcal{T}^*_n(\mu_1)\otimes I)
\right)
\label{eq:new3}\\
&&\quad\times
\left(\omega _1(\mu_{12})
\mathcal{X}^*_
{1,n}(\mu_1,\mu_2|\la_1,\cdots,\la_n)
+\omega _2(\mu_{12})
\mathcal{X}^*
_{2,n}(\mu_1,\mu_2|\la_1,\cdots,\la_n)\right)
\nn,
\ena
with
\begin{align}
\mathcal{T}^*_n(\mu)
(\mathcal{O})=L_{1}(\mu-\la_1)^{-1}\cdots L_{ n}(\mu-\la_n)^{-1}
\ \mathcal{O}\ L_n(\mu-\la_n)\cdots L_1(\mu-\la_1),
\nn
\end{align}
and 
$\mathcal{X}^*_{1,n}$, $\mathcal{X}^*_{2,n}$ 
are constructed from $\mathcal{T}_n^*(\mu)$ as before. 

Let us discuss the last formula briefly. 
Consider the operators $\mathcal{O}$ which act as identity 
either on the last or on the first site
(i.e., $\mathcal{O}=\mathcal{O}'\otimes I$ or 
$\mathcal{O}=I\otimes \mathcal{O}'$). 
For such operators, 
we have 
\begin{align}
&\Omega ^*_n(I\otimes\mathcal{O}')
=I\otimes \Omega ^*_{n-1}(\mathcal{O}'),
\label{contr21}\\
&\Omega ^*_n(\mathcal{O}'\otimes I)
=\Omega ^*_{n-1}(\mathcal{O}')\otimes I.
\label{contr22}
\end{align}
Eq. \eqref{contr22} is obvious from the definition, while 
\eqref{contr21} is non-trivial and follows from 
\eqref{eq:contraction}. 
This motivates us 
to consider a `universal' operator 
\begin{align}
&\mathcal{T}^*(\mu)(\mathcal{O})
\nn\\
&=
\lim _{N\to \infty}L_{-N}(\mu-\la_{-N})^{-1}
\cdots L_{N}(\mu-\la_N)^{-1}
\ \mathcal{O}\ L_N(\mu-\la_N)\cdots L_{-N}(\mu-\la_{-N}).
\nn
\end{align}
Introducing further the normalized trace
$$
\mathbf{tr}=\lim _{N\to \infty}\(\frac 1 2 \text{tr}_{V_{-N}}\cdots
\frac 1 2 \text{tr}_{V_{N}}\),
$$
and defining $\Omega^*$ using $\mathcal{T}^*(\mu)$, 
we can write down a universal formula
\bea
\langle \text{vac}|\mathcal{O}|\text{vac}\rangle=
\mathbf{tr}\(e^{\Omega ^*}(\mathcal{O})\), 
\label{eq:new4}
\ena
which does not refer to the size $n$ of the subsystem.
For any operator acting on a finite sublattice, 
the right hand side reduces automatically to 
this sublattice due to (\ref{contr21})--(\ref{contr22}). 
Consider the action of the universal 
$\mathcal{T}^*(\mu)$ on a given operator $\mathcal{O}$. 
If $\mathcal{O}$ acts on sites $1,\cdots,n$, then  
the infinite right tail of $L$-operators $L_j$ with $j>n$ cancels. 
However, the left tail with $j<1$ remains. 
In integrable quantum field theory, 
this situation is typical for the action of non-local charges which 
transform local operators into non-local ones with an 
infinite tail in one direction \cite{LS}. 
Nevertheless, a beautiful feature of our 
construction is that, 
when we substitute $\mathcal{T}^*(\mu)$ 
into the trace and integrate, we obtain the operator 
$\Omega ^*$ which sends a local operator to a local one  
because of (\ref{contr22}). 

In our opinion, this is the most important
property of our construction which deserves further understanding.

\section{Invariant operators}\label{sec:invariant}

As we have seen in the previous section, 
$U'_q(\slth)$ operates on our space of local operators \eqref{eq:local}. 
The algebra $U'_q(\slth)$ contains two subalgebras 
isomorphic to $U_q(\slt)$, one generated by $e_0,f_0,q^{\pm h_0/2}$ and the other 
by $e_1,f_1,q^{\pm h_1/2}$. 
In this section, we consider the subspace of local operators 
which are invariant under one of these subalgebras, 
and show that their correlation functions 
do not contain the transcendental function $\omega _2(\la)$. 
To fix the idea, let us choose the subalgebra generated by 
$e_0,f_0,q^{\pm h_0/2}$ and set 
\be
\mathcal{L}^{\rm inv}_n
&=&
\{\mathcal{O}\in\mathcal{L}_n\mid
x\cdot\mathcal{O}=\mathcal{O}\cdot x\quad
(x=e_0,f_0,q^{h_0/2})\}. 
\en

In the present context, it is more convenient to use 
the formula for $\Omega^*_n$ using the invariant trace 
(see the end of Section \ref{sec:XXZansatz}), 
\bea
&&\Omega^*_n(\la_1,\cdots,\la_n)
=\frac{1}{2\kappa^2}\int\!\!\int
\frac{d\mu_1}{2\pi i}\frac{d\mu_2}{2\pi i}
{\rm Tr}_{2,2}\left(B^q(\mu_{1,2})
(I\otimes \mathcal{T}^*_n(\mu_2))
(\mathcal{T}^*_n(\mu_1)\otimes I)
\right)
\label{eq:new5}\\
&&\quad\times
\left(\omega _1(\mu_{12})
\mathcal{X}^{q*}_
{1,n}(\mu_1,\mu_2|\la_1,\cdots,\la_n)
+\omega _2(\mu_{12})
\mathcal{X}^{q*}_{2,n}(\mu_1,\mu_2|\la_1,\cdots,\la_n)\right)
\nn.
\ena
As before, $\mathcal{X}^{q*}_{1,n}$, $\mathcal{X}^{q*}_{2,n}$ 
are constructed from $\mathcal{T}_n^*(\mu)$ 
using ${\rm Tr}^q_\mu$. 

\begin{lem}
The space $\mathcal{L}^{\rm inv}_n$ is invariant under the 
operator $\Omega^*_n(\la_1,\cdots,\la_n)$. 
\end{lem}
\begin{proof}
First we show that ${\rm Tr}^q_{\mu}\mathcal{T}_n^*(\mu)$ 
preserves the space $\mathcal{L}_n^{\rm inv}$. 
Abbreviating arguments, we write 
\be
\mathcal{T}_n^*(\mathcal{O})=T^{-1}(1\otimes\mathcal{O}) T, 
\en
where $T=L_n\cdots L_1$,  
$L_j=L_j(\mu-\la_j)$ and $\mu=(\mu_1+\mu_2)/2$. 

Until the end of the proof, we let 
$x$ stand for $e_0,f_0,q^{h_0/2}$. 
The operator $T$ belongs to $U_q(\slth)\otimes\End(W)$ where 
$W=V(\lambda_1)\otimes\cdots\otimes V(\lambda_n)$. 
It satisfies the intertwining property
\be
\sum T(x'_i\otimes x''_i)=\sum(x''_i\otimes x'_i)T\,,
\en
where 
$\Delta(x)=\sum x'_i\otimes x''_i$.  
This equation can be rewritten as
\be
T(1\otimes x)=\sum(x'''_i\otimes x''_i)T(S^{-1}(x'_i)\otimes1),
\en
where we have set 
$(\Delta\otimes I)\circ\Delta(x)=\sum
x'_i\otimes x''_i\otimes x'''_i$.  
Using the invariance of $\mathcal O$
and the intertwining property again, we obtain, 
in the notation above, 
\bea
T^{-1}(1\otimes\mathcal{O})T(1\otimes x)
=
\sum(x''_i\otimes x'''_i)
T^{-1}(1\otimes\mathcal{O})T(S^{-1}(x'_i)\otimes1).
\label{eq:Ti}
\ena
We have $q^{-h_0}xq^{h_0}=S^2(x)$ and $\pi_\la(q^{-h_0})=q^{H}$,  
from which follows the invariance property
\be
{\rm Tr}_\lambda^q(\ad'x\,A)=\e(x){\rm Tr}_\lambda^q(A)\,.
\en
Taking $\Tr^q_\mu$ of both sides of \eqref{eq:Ti} and using
the above invariance, we find that 
$$
x\cdot{\rm Tr}^q_{\mu}\mathcal{T}^*(\mathcal{O})=
{\rm Tr}^q_{\mu}\mathcal{T}^*(\mathcal{O})\cdot x.  
$$

Let us show that the operator 
\bea
{\rm Tr}_{2,2}\left(B^q(\mu_{1,2})
(I\otimes\mathcal{T}^*_n(\mu_2))(\mathcal{T}^*_n(\mu_1)\otimes I)\right)
\label{eq:BTT}
\ena
also preserves $\mathcal{L}_n^{\rm inv}$. 
The relevant operators act on the tensor product
$\C^2\otimes\C^2\otimes W$. 
To unburden the notation, let us write 
$T_{12}=(T(\mu_1)\otimes I)(I\otimes T(\mu_2))$, 
$B^q_{12}=B^q(\mu_{12})$ and $\mathcal{O}_3=\mathcal{O}$, 
indicating the tensor components by the suffix. 
Thus the action of \eqref{eq:BTT} on $\mathcal{O}$ is 
${\rm Tr}_{2,2}
\left(B^q_{12}T_{12}^{-1}\mathcal{O}_3 T_{12}\right)$. 

Writing again $\Delta(x)=\sum x'_i\otimes x''_i$,
we have 
\bea
\sum (x'_i)_1(x''_i)_2 B^q_{12}
=\sum B^q_{12}(x'_i)_1 (x''_i)_2
=\epsilon(x)B^q_{1,2}.
\label{eq:Be}
\ena
Using \eqref{eq:Be} together with 
the intertwining property of $T$ and 
the invariance of $\mathcal O$, we find 
\be
&&B^q_{1,2}T_{12}^{-1}\mathcal{O}_3T_{12} x_3 
=\sum B^q_{1,2}(x_i'')_3T_{12}^{-1}\mathcal{O}_3 T_{12}
\Delta(S^{-1}(x'_i))_{12}.
\en
Taking trace, moving the last factor 
by cyclicity and using \eqref{eq:Be} again, we obtain 
\be
&&{\rm Tr}_{2,2}\left(B^q(\mu_{1,2})
(I\otimes\mathcal{T}^*_n(\mu_2))
(\mathcal{T}^*_n(\mu_1)\otimes I)\right)(\mathcal{O})
\cdot x 
\\
&&\quad=\sum{\rm Tr}_{2,2}
\bigl(\Delta(S^{-1}(x'_i))_{12}(x''_i)_3
B^q_{12}T_{12}^{-1}\mathcal{O}_3T_{12}\bigr)
\\
&&\quad=x\cdot 
{\rm Tr}_{2,2}\left(B^q(\mu_{1,2})
(I\otimes\mathcal{T}^*_n(\mu_2))
(\mathcal{T}^*_n(\mu_1)\otimes I)\right)(\mathcal{O}),
\en
which was to be shown. 
\end{proof}

\begin{lem}
For an invariant operator 
$\mathcal{O}\in\mathcal{L}_n^{\rm inv}$, 
we have $\mathcal{X}^{q*}_{2,n}(\mu_1,\mu_2)(\mathcal{O})=0$. 
\end{lem}
\begin{proof}
Lemma means that  the trace
\be
&&{\rm Tr}_{\mu_{12}}^q
(L_1(\mu-\lambda_1)^{-1}
\cdots L_n(\mu-\lambda_n)^{-1}
\mathcal{O}
L_n(\mu-\lambda_n)\cdots L_1(\mu-\lambda_1))
\\
&&={\rm Tr}_{\mu_{12}}^q
(L_1(\lambda_1-\mu)\cdots L_n(\lambda_n-\mu)
\mathcal{O}
L_n(\lambda_n-\mu)^{-1}\cdots L_1(\lambda_1-\mu)^{-1})
\en
does not produce terms proportional to $\mu_{12}$. 
We prove this assertion 
by passing to the vector language
and performing a gauge transformation. 

Under the isomorphism \eqref{eq:local}, 
an invariant operator $\mathcal{O}$ is sent to a vector 
$v\in V(\la_1)\otimes\cdots\otimes V(\la_n)\otimes
V(\la_n-1)\otimes\cdots\otimes V(\la_1-1)$ 
invariant under the action of $e_0,f_0,q^{h_0/2}$. 
Set $g=q^{(1/2)\sum_{j=1}^n
(\la_j\sigma^3_j+(\la_j-1)\sigma^3_{\bar j})}$, 
and introduce the gauge transformation 
\be
&&e=gf_0g^{-1}, \quad f=ge_0g^{-1},\quad q^{h/2}=q^{-h_0/2},
\\
&&\ell'(\la)=q^{(\la/2)\sigma^3}\ell(\la)q^{-(\la/2)\sigma^3}.
\en
Then $v'=gv$ belongs to the subspace 
$\left(V^{\otimes 2n}\right)^{\rm inv}$
of vectors invariant under $U_q(\slt)$ generated by $e,f,q^{h/2}$.

For the proof, we show the following slightly more general statement:
for any $v'\in \left(V^{\otimes 2n}\right)^{\rm inv}$
and $\la_1,\cdots,\la_{2n}$, 
\be
{\rm Tr}_\mu^q\left(\ell'_1(\la_1)\cdots\ell'_{2n}(\la_{2n})
\right)v'
\en
belongs to 
$\C[q^{\pm \mu},q^{\pm \la_1},\cdots,q^{\pm \la_{2n}}]$.

First consider the case $n=1$,  
choosing $v'$ to be $s^q=qv_+\otimes v_--v_-\otimes v_+
\in (V^{\otimes 2})^{\rm inv}$. 
Note that 
\be
&&{\rm Tr}_\mu^q(FEA) 
={\rm Tr}_\mu^q\left(
\left[\frac{\mu+1+H}{2}\right]
\left[\frac{\mu-1-H}{2}\right]
A\right), 
\\ 
&&{\rm Tr}_\mu^q(EFA) 
={\rm Tr}_\mu^q\left(
\left[\frac{\mu+1-H}{2}\right]
\left[\frac{\mu-1+H}{2}\right]
A\right).  
\en 
Direct computation using these relations 
shows that the entries of 
$\ell'_1(\la_1)\ell'_2(\la_2)s^q_{12}$
can be reduced to elements of the subalgebra generated by 
$q^{-H},Eq^{-H/2},Fq^{-H/2}$.
Since $q^H$ does not appear, ${\rm Tr}_\mu^q$ does not
produce terms proportional to $\mu$. 

In the general case, the same argument shows that  
the assertion holds for 
$v'={\bf s}^q=s^q_{12}\cdots s^q_{2n-1\,2n}$.
From the Yang-Baxter relation, 
the same is true for vectors obtained 
from ${\bf s}^q$ by acting with an 
arbitrary number of matrices 
$\check{r}'_{i\,i+1}(\la_{i\,i+1})$, 
where
\be
\check{r}'_{i\,i+1}(\la)=P_{i\,i+1}
q^{\la\sigma_i/2}r_{i\,i+1}(\la)q^{-\la\sigma_i/2}.
\en
The operators $\check{r}'_{i\,i+1}(\la)$ 
are linear combinations 
of $1$ and the 
generators $e_i=\check{r}'_{i\,i+1}(-1)$ ($i=1,\cdots,2n-1$)
of the Temperley-Lieb algebra,  
and vice versa. 
It is well known that the space 
$\left(V^{\otimes 2n}\right)^{\rm inv}$
is generated from ${\bf s}^q$ by the action of the
Temperley-Lieb algebra.  
Hence the assertion is true for all 
$v'\in\left(V^{\otimes 2n}\right)^{\rm inv}$. 
\end{proof}

In summary, let us present the final result for 
invariant operators. 
In the notation of section \ref{sec:alternative}, we have
\begin{thm}
Consider a local operator $\mathcal{O} $ invariant under the
action of $U_q(\slt)$ generated by $e_0,f_0, q^{h_0/2}$. 
For such an operator, we have 
\bea
\langle \text{vac}|\mathcal{O}|\text{vac}\rangle=
\mathbf{tr}\(e^{\Omega _{\text{inv}}^*}(\mathcal{O})\)
\label{eq:new2inv}
\ena
where
\bea
&&\Omega^*_{\text{inv}}
=\frac{1}{2\kappa^2}\int\!\!\int\frac{d\mu_1}{2\pi i}\frac{d\mu_2}{2\pi i}
\label{eq:newinv}\\
&&\quad\times
\omega _1(\mu_{12})
{\rm Tr}_{2,2}\left(B^q(\mu_{1,2})\mathcal{T}_n^{*}(\mu_2)
\otimes \mathcal{T}_n^{*}(\mu_1)
\right)
\Tr^q_{\mu_{1,2}}\mathcal{T}_n^{*}\(
\frac {\mu _1+\mu _2} 2\).
\nn
\ena
\end{thm}
Formula \eqref{eq:newinv} 
is quite similar to the one in the 
XXX case \cite{BJMST4}.

Various methods are known for constructing 
bases of invariant operators (see e.g. \cite{PM}). 
For example, a basis for $n=3$ is 
$I$, $U_{12}$, $U_{23}$, $U_{12}U_{23}$, $U_{23}U_{12}$.  
Here the operators $U_{i\,i+1}$, after the gauge transformation
${U}_{i\,i+1}'=g'{U}_{i\,i+1} g^{'-1}$ by 
$g'=q^{(1/2)\sum_{j=1}^n\lambda_j\sigma^3_j}$,  
are the (negative of the)  
generators of the Temperley-Lieb algebra,   
$$
U'_{i\,i+1}=
\frac{1}{2}
\left(
\sigma ^1_i\sigma ^1_{i+1}
+\sigma ^2_{i}\sigma ^2_{i+1}
+\frac {q+q^{-1}} 2\bigl(\sigma ^3_i\sigma ^3_{i+1}-1\bigr)
+\frac {q-q^{-1}} 2\bigl(-\sigma _i^3+\sigma ^3_{i+1}\bigr)
\right).  
$$
(It is nothing but the local density 
of the Pasquier-Saleur Hamiltonian, see below).  
In the homogeneous case, their expected values are 
\be
&&\langle{\rm vac}|U_{i,i+1}|{\rm vac}\rangle
=-\kappa\,a_0,
\\
&&
\langle{\rm vac}|U_{12}U_{23}|{\rm vac}\rangle
=\langle{\rm vac}|U_{23}U_{12}|{\rm vac}\rangle
=-\frac{1}{\cos\pi\nu}
(\kappa a_0-3\kappa^3 a_2),
\en
where $\kappa=\sin\pi\nu/(\pi\nu)$ and  
$$
a_{m}=
\int_{-\infty}^\infty t^{m}
\frac{\sinh (1/\nu-1)t}{\sinh (t/\nu) \cosh t}\,dt. 
$$

Let us explain the physical meaning of correlation functions of invariant operators. We shall consider the case of massless regime
$q=e^{\pi i \nu}$ ($0<\nu<1$), 
because in the present context it is more interesting 
from the point of view of physics. 
It is well known that in the continuous limit
the massless XXZ model is described by 
CFT with the central charge $c=1$.  
In the continuous field theory, 
starting with CFT with $c=1$
one can obtain CFT with $c=1-\frac {6\nu ^2}{1-\nu}$ 
by modifying the energy-momentum tensor 
and introducing screening operators.  
There is a construction which gives a lattice version of this
procedure, and which is closely related to the invariance
under the quantum group. 

Consider first the model in the finite volume. The XXZ Hamiltonian
with the usual periodic boundary condition 
$$ 
H_L =\frac{1}{2}
\sum\limits _{j=1}^L\(\sigma ^1_j \sigma^1_{j+1}+
\sigma ^2_j \sigma^2_{j+1}+
\cos(\pi\nu)\sigma ^3_j \sigma^3_{j+1}\),
\quad \sigma ^a_{L+1}=\sigma ^a_1, 
$$
is not invariant under
the quantum group. 
However, following Pasquier and Saleur \cite{PS}
one can introduce an invariant Hamiltonian with specific boundary
conditions:
$$ 
H_L^{\text{inv}} 
=\frac{1}{2}\sum\limits _{j=1}^{L-1}
\(\sigma ^1_j \sigma^1_{j+1}+
\sigma ^2_j \sigma^2_{j+1}+\cos(\pi\nu)\sigma ^3_j \sigma^3_{j+1}\) + 
\frac{1}{2}i\sin(\pi\nu)(\sigma ^3_L-\sigma ^3_1).  
$$
In the infinite volume limit,  
this Hamiltonian has CFT with 
$c=1-\frac {6\nu ^2}{1-\nu}$ as the continuous limit. 
In the finite volume there are 
significant differences between $H_L$ and $H_L^{\text{inv}}$;
for example, Bethe Ansatz equations are very different. 
However, following the general logic 
(see, for example, \cite{RS}) we believe that 
in the infinite volume the ground state of the 
invariant model is obtained by projection 
of the original ground state onto the invariant 
subspace:
$$
|\text{vac}\rangle_{\text{inv}}=\cP _{\text{inv}}|
\text{vac}\rangle,  
$$
where $\cP _{\text{inv}}$ denotes the 
projection operator.
Then the correlation function of any operator in the 
invariant model coincides with that  
of an invariant operator in the original model:
$$
{\ }_{\text{inv}}\langle \text{vac}|\mathcal{O}|\text{vac}\rangle_{\text{inv}}=
\langle \text{vac}|\(\cP _{\text{inv}}\mathcal{O}\cP _{\text{inv}}\)|\text{vac}\rangle. 
$$
So, the correlation functions considered 
in this section 
describe the lattice version of CFT with 
$c=1-\frac {6\nu ^2}{1-\nu}$.
The fact that they can be expressed 
in terms of a single transcendental
function was predicted in \cite{BKS}. 
Certainly, the most interesting
question is that of rational $\nu$ when 
the space of local operators is restricted. 
We shall consider this situation in the future.

\section{XYZ model}\label{sec:XYZ}

A considerable part of the previous sections 
can be generalized to the case of the XYZ chain
 \begin{eqnarray}
H_{\rm XYZ}=\frac{1}{2}
\sum_{k=-\infty}^{\infty}
\left( 
I_1\sigma_{k}^1\sigma_{k+1}^1+
I_2\sigma_{k}^2\sigma_{k+1}^2+
I_3\sigma_{k}^3\sigma_{k+1}^3
\right).  
\label{eq:XYZ}
\end{eqnarray}
In \cite{BJMST3} we put forward a conjectural formula 
for the density matrix in the elliptic setting.  
In comparison with the XXX and XXZ chains, 
however, the results obtained are incomplete: 
we have so far  been unable to verify that the 
expression written down in 
\cite{BJMST3} satisfies the reduced qKZ equation. 
In this section, we rewrite the formula conjectured in 
\cite{BJMST3} into a form close to 
\eqref{eq:ansatz}--\eqref{eq:newformula}. 

In the following, we denote by $\theta_a(t)$ ($a=1,2,3,4$) 
the Jacobi elliptic theta function with modulus $\tau$. 
We fix a generic complex number $\eta$ and   
use the scaled spectral parameter $t=\la \eta$. 
We deal with functions which have period $1$ 
in the variable $t$. 
The parameters in the Hamiltonian \eqref{eq:XYZ} are given by 
$(1/2)I_a=\theta_{a+1}(2\eta)/\theta_{a+1}(0)$. 

In the XYZ chain, the role of $U_q(\slt)$ in the XXZ chain 
is played by the Sklyanin algebra \cite{Sk}.  
It is an associative algebra $\cA$ generated by four symbols  
$S_\alpha$ ($\alpha=0,1,2,3$) 
satisfying the defining quadratic relations 
\bea
&&[S_0,S_a]=iJ_{bc}(S_bS_c+S_cS_b), 
\label{eq:quad1}\\
&&[S_b,S_c]=i(S_0S_a+S_aS_0).  
\label{eq:quad2}
\ena
Here $(a,b,c)$ runs over cyclic permutations of $(1,2,3)$. 
The structure constants $J_{bc}=-(J_b-J_c)/J_a$ 
are parametrized as 
\be
&&J_a=\frac{\theta_{a+1}(2\eta)\theta_{a+1}(0)}
{\theta_{a+1}(\eta)^2}\,.
\en
The algebra $\cA$ has two basic central elements, 
$K_0=\sum_{\alpha=0}^3S_\alpha^2$ and 
$K_2=\sum_{a=1}^3 J_aS_a^2$. 
We will consider only representations of $\cA$ 
on which these central elements act as scalars, 
\bea
&&
K_0=4\Bigl[\frac{d}{2}\Bigr]^2,
\quad K_2=4\Bigl[\frac{d+1}{2}\Bigr]\Bigl[\frac{d-1}{2}\Bigr].
\label{eq:Casimir}
\ena
Here and after we set 
\be
[t]=\frac{\theta_1(2t)}{\theta_1(2\eta)}.
\en 
The parameter $d$ plays the role of the `dimension'. 
For each non-negative integer $k$, the Sklyanin algebra 
possesses an analog $\pi^{(k)}$ 
of the $(k+1)$-dimensional irreducible representation of $\slt$,
on which the above relations are valid with $d=k+1$.   

The $L$ operator associated with the XYZ chain has the following form. 
\be
&&L(t):=\frac{\rho(t,d)}{[t+(d/2)\eta]}\,\ell(t)\,,
\\
&&\ell(t)=\frac{1}{2}\sum_{\alpha=0}^3
\frac{\theta_{\alpha+1}(2t+\eta)}{\theta_{\alpha+1}(\eta)}
S_\alpha\otimes \sigma^\alpha\,.
\en
Here $\rho(t,d)$ is a normalization factor 
(see \eqref{eq:rho_ell}) which satisfies 
\be
&&\rho(t,d)\rho(-t,d)=1,\\
&&\frac{\rho(t,d)}{[t+(d/2)\eta]}
\frac{\rho(t-\eta,d)}{[t-\eta+(d/2)\eta]}
=\frac{1}{[(d/2)\eta-t][(d/2)\eta+t]}.
\en
We define the monodromy matrix by the same formula \eqref{eq:T}. 

The $R$ matrix $R(t)$ for the XYZ chain is defined by the above 
formula with $d=2$ and $S_\alpha$ being represented by 
$\sigma^\alpha$. 
We need the following three functions
\footnote{Note that $\omega_{i}$'s here are 
those in \cite{BJMST3} multiplied by $\kappa/4$.}
$\omega_i=\omega_i(t,\eta,\tau)$ 
which enter the formula for $h_n$: 
\bea
&&\kappa d\log\varphi =\omega_1dt+\omega_2d\eta+\omega_3d\tau, 
\label{eq:om}
\ena
where 
\be
&&\kappa=\frac{\theta_1(2\eta)}{2\theta'_1(0)},
\\
&&\varphi(t):=\rho(t,2)\cdot 
\left(\frac{[\eta-t]}{[\eta+t]}\right)^{1/4}.
\en

In the XYZ case, the trace functional is defined as follows. 
For each element $A$ of the Sklyanin algebra, there exists a unique entire 
function ${\rm Tr}_\lambda A$ with the properties \cite{BJMST3} 
\begin{enumerate}
\item $\Tr_\la A\bigl|_{\la=d}=\tr\, \pi^{(d)}(A)$ 
holds for all positive integers $d$, 
\item If $A$ is a monomial in the $S_\alpha$ 
of homogeneous degree $n$,  $\Tr_\la A$ has the form
\bea
\Tr_{\frac{t}{\eta}} A=
\theta_1(t)^n\times 
\begin{cases}
g_{A,0}(t) & (\mbox{ $n$: odd }),\\
g_{A,1}(t)-\frac{t}{\eta} g_{A,2}(t) & (\mbox{ $n$: even }),\\
\end{cases}
\label{eq:trdec}
\ena
where 
$g_{A,0}(t)$, $g_{A,2}(t)$ and 
$g_{A,3}(t):=g_{A,1}(t+\tau)-g_{A,1}(t)$ 
are elliptic functions with periods $1,\tau$. 
In addition, $g_{A,1}(t+1)=g_{A,1}(t)$. 
\end{enumerate}

Let us introduce $X_{a,n}$ ($a=1,2,3$) by 
\be
&&
{\rm Tr}_{s_{12}/\eta}\left(T\Bigl(\frac{s_1+s_2}{2}\Bigr)\right)
=
X_{1,n}(s_1,s_2|t_1,\cdots,t_n)
-\frac{s_{12}}{\eta}X_{2,n}(s_1,s_2|t_1,\cdots,t_n),
\\
&&X_{3,n}(s_1,s_2|t_1,\cdots,t_n)=
X_{1,n}(s_1+\tau,s_2|t_1,\cdots,t_n)-
X_{1,n}(s_1,s_2|t_1,\cdots,t_n)\,,
\en
where $X_{a,n}$ ($a=1,2,3$) have the following periodicity.
\be
&&
X_{a,n}(s_1+1,s_2|t_1,\cdots,t_n)
=X_{a,n}(s_1,s_2|t_1,\cdots,t_n)
\quad (a=1,2,3),
\\
&&
X_{a,n}(s_1+\tau,s_2|t_1,\cdots,t_n)=
X_{a,n}(s_1,s_2|t_1,\cdots,t_n)
\quad (a=2,3). 
\en

\begin{prop}
The conjectural formula in \cite{BJMST3} for $h_n$ can be written as 
\bea
&&\Omega_n(t_1,\cdots,t_n)
=
\frac{(-1)^n}{2\kappa^2}
\int\!\!\int 
\frac{ds_1}{2\pi i}\frac{ds_2}{2\pi i}
\left(\sum_{a=1}^3\omega_a(s_{12})
X_{a,n}(s_1,s_2|t_1,\cdots,t_n)
\right)
\label{eq:Omega_ell}\\
&&
\quad\times
{\rm Tr}_{2,2}
\left(T_n(s_1)\otimes T_n(s_2)\cdot B(s_{1,2})\right),
\nonumber
\ena
where 
\be
B(t)=-\frac{1}{4}
\frac{[t][2\eta]}{[t+\eta][t-\eta]}
\sum_{a=1}^3\frac{\theta_{a+1}(2t)}{\theta_{a+1}(2\eta)}
\sigma^a\otimes\sigma^a\,.
\en
\end{prop}
The rewriting procedure is sketched
at the end of Appendix \ref{app:new}.

\appendix

\section{Connection to previous results}\label{sec:review}

Here we give the details about the derivation of the formula 
\eqref{eq:newformula} and 
the second one using \eqref{eq:GGt2} and \eqref{eq:Bsigma}. 
%\eqref{eq:def-omega-gauge}. 
First, in subsection \ref{app:a-1}, 
we recall the previous result in \cite{BJMST2}, where 
an algebraic formula for a solution 
of the reduced qKZ equation is constructed. 
Next we rewrite the formula into 
the exponential form in subsection \ref{app:a-2} 
(see Theorem \ref{thm:app-a-1}). 
Finally we obtain the integral formula 
\eqref{eq:newformula} and 
the second one using \eqref{eq:GGt2} and \eqref{eq:Bsigma}
in subsection \ref{app:new}. 
The above rewriting procedure is applicable also to the elliptic case. 
We discuss it briefly at the end of subsection \ref{app:new}. 

\subsection{Algebraic construction of a solution to the reduced 
qKZ equation}\label{app:a-1}

In \cite{BJMST2} a solution $\{h_{n}\}_{n=0}^{\infty}$ of 
the equations \eqref{eq:rqkz1}--\eqref{eq:rqkz3} is 
constructed in an algebraic way. 
Let us recall it here. 

First we define the operator 
\begin{eqnarray*}
{}_{n}X_{n-2}^{(i, j)}(\lambda_{1}, \ldots , \lambda_{n}) 
\in {\rm End}(V^{\otimes 2(n-2)}, V^{\otimes 2n}) 
\end{eqnarray*}
for $1 \le i<j \le n$ as follows. 
Set 
\begin{eqnarray}
&& 
{}_{n}X_{n-2}(\lambda_{1}, \ldots , \lambda_{n})(u):=
\frac{1}{[\lambda_{1, 2}]\prod_{p=3}^{n}[\lambda_{1,p}][\lambda_{2,p}]}
\label{eq:def-X} \\
&& \hspace{8em} {}\times 
{\rm Tr}_{\lambda_{1, 2}}\left(
t^{[1]}_{n}\bigl(\frac{\lambda_{1}+\lambda_{2}}{2};
             \lambda_{1}, \ldots , \lambda_{n}\bigr) \right)
(s_{1, \bar{2}}s_{\bar{1}, 2}
 u_{3, \ldots , n, \bar{n}, \ldots ,\bar{3}}), 
\nonumber  
\end{eqnarray}
where 
\begin{eqnarray*}
t_{n}^{[i]}(\lambda;\lambda_{1}, \ldots , \lambda_{n})&:=&
\ell_{\bar{1}}(\lambda-\lambda_{1}-1) \cdots 
\widehat{\ell_{\bar{i}}(\lambda-\lambda_{i}-1)} \cdots 
\ell_{\bar{n}}(\lambda-\lambda_{n}-1) \\ 
&\times& 
\ell_{n}(\lambda-\lambda_{n}) \cdots 
\widehat{\ell_{i}(\lambda-\lambda_{i})} \cdots 
\ell_{1}(\lambda-\lambda_{1}).
\end{eqnarray*}
Then ${}_{n}X_{n-2}^{(i,j)}$ is defined by 
\begin{eqnarray*}
&& 
{}_{n}X_{n-2}^{(i,j)}(\lambda_{1}, \ldots , \lambda_{n}) \\ 
&& {}:=
\overleftarrow{\mathbb{R}}_{n}^{(i, j)}(\lambda_{1}, \ldots , \lambda_{n})\,
\overleftarrow{P}_{n}^{(i,j)}
\cdot 
{}_{n}X_{n-2}(\lambda_{i}, \lambda_{j}, \lambda_{1}, \ldots , 
 \widehat{\lambda_{i}}, \ldots , \widehat{\lambda_{j}}, \ldots , 
 \lambda_{n}), 
\end{eqnarray*}
where  %\mathbb{R}_{n}^{(i, j)}(\lambda_{1}, \ldots , \lambda_{n}) \cdot 
\begin{eqnarray*}
\overleftarrow{\mathbb{R}}_{n}^{(i, j)}(\lambda_{1}, \ldots , \lambda_{n})
&:=&
R_{i, i-1}(\lambda_{i, i-1}) \cdots 
R_{i,1}(\lambda_{i,1}) \\ 
&\times& 
R_{j, j-1}(\lambda_{j, j-1}) \cdots 
\widehat{R_{j, i}(\lambda_{j, i})} \cdots 
R_{j,1}(\lambda_{j, 1}) \\ 
&\times& 
R_{\overline{i-1}, \overline{i}}(\lambda_{i-1, i}) \cdots 
R_{\overline{1}, \overline{i}}(\lambda_{1,i}) \\
&\times& 
R_{\overline{j-1}, \overline{j}}(\lambda_{j-1, j}) \cdots 
\widehat{R_{\overline{i}, \overline{j}}(\lambda_{i, j})} \cdots 
R_{\overline{1}, \overline{j}}(\lambda_{1, j})
\end{eqnarray*}
and 
\begin{eqnarray*}
\overleftarrow{P}_{n}^{(i,j)}:=P_{i, i-1} \cdots P_{2,1} \cdot 
P_{j, j-1} \cdots P_{3,2} \cdot 
P_{\overline{i-1}, \overline{i}} \cdots P_{\overline{1}, \overline{2}}
\cdot 
P_{\overline{j-1}, \overline{j}} \cdots P_{\overline{2}, \overline{3}}.
\end{eqnarray*}
Note that the operator 
$\overleftarrow{\mathbb{R}}_{n}^{(i, j)}
 (\lambda_{1}, \ldots , \lambda_{n})$ 
is a rational function in 
$\zeta_{j}=e^{\pi i \nu \lambda_{j}} \, (j=1, \ldots , n)$ 
%$\zeta_{1}, \ldots , \zeta_{n}$,  
%where $\zeta_{j}=e^{\pi i \nu \lambda_{j}}$,  
because $\rho(\lambda)\rho(-\lambda)=1$.  

{}From the definition \eqref{eq:trfcn} of ${\rm Tr}_{\lambda}$, 
the operator ${}_{n}X_{n-2}^{(i, j)}$ can be written uniquely 
in the following form:
\begin{eqnarray}
{}_{n}X_{n-2}^{(i, j)}(\lambda_{1}, \ldots , \lambda_{n})=
{}_{n}G_{n-2}^{(i, j)}(\zeta_{1}, \ldots , \zeta_{n})
-\lambda_{i,j}\cdot 
{}_{n}\tilde{G}_{n-2}^{(i, j)}(\zeta_{1}, \ldots , \zeta_{n}),
\label{eq:defGG}
\end{eqnarray}
where 
${}_{n}{G}_{n-2}^{(i, j)}(\zeta_{1}, \ldots , \zeta_{n})$ 
and 
${}_{n}\tilde{G}_{n-2}^{(i, j)}(\zeta_{1}, \ldots , \zeta_{n})$ 
are rational functions in  
$\zeta_{1}, \ldots , \zeta_{n}$. 
%$\zeta_{j}=e^{\pi i \nu \lambda_{j}} \, (j=1, \ldots , n)$. 
Take some meromorphic functions 
$\omega_{j}(\lambda) \, (j=1,2)$ and 
consider the operator 
\begin{eqnarray*}
{}_{n}\Omega_{n-2}^{(i,j)}(\lambda_{1}, \ldots , \lambda_{n}):=
\omega_{1}(\lambda_{i,j})\cdot 
{}_{n}G_{n-2}^{(i,j)}(\zeta_{1}, \ldots , \zeta_{n})+
\omega_{2}(\lambda_{i,j})\cdot 
{}_{n}\tilde{G}_{n-2}^{(i,j)}(\zeta_{1}, \ldots , \zeta_{n}).
\end{eqnarray*}
For an ordered set of indices $K=\{k_{1}, \ldots, k_{m}\}$ 
$(1 \le k_{1} < \cdots < k_{m} \le n)$, we use the abbreviation 
\begin{eqnarray*}
\Omega_{K, (k_{i}, k_{j})}={}_{m}\Omega_{m-2}^{(i,j)}
(\lambda_{k_{1}}, \ldots , \lambda_{k_{m}}). 
\end{eqnarray*}
Define 
\begin{eqnarray}
&& 
h_{n}(\lambda_{1}, \ldots , \lambda_{n}) 
\label{eq:original-solution-1} \\ 
&& {}:=
\sum_{m=0}^{[n/2]}\frac{(-1)^{m}}{2^{n-2m}}
\sum
\Omega_{K_{1}, (i_{1}, j_{1})} \cdot 
\Omega_{K_{2}, (i_{2}, j_{2})} \cdot \cdots 
\Omega_{K_{m}, (i_{m}, j_{m})} ({\bf s}_{n-2m}).
\nonumber
\end{eqnarray}
Here 
\begin{eqnarray*}
{\bf s}_{m}:=\prod_{p=1}^{m} s_{p, \bar{p}} \in V^{\otimes 2m} 
\end{eqnarray*}
and the second sum in \eqref{eq:original-solution-1} 
is taken over all sequences 
$i_{1}, \ldots , i_{m}, j_{1}, \ldots , j_{m}$ 
of distinct elements of $K_{1}=\{1, \ldots , n\}$ such that 
\begin{eqnarray*}
i_{1}< \cdots < i_{m}, \quad i_{1}<j_{1}, \quad \cdots , \quad  
i_{m}<j_{m}. 
\end{eqnarray*}
The sets $K_{2}, \ldots , K_{m}$ are defined by 
$K_{l}=K_{l-1}\setminus\{i_{l}, j_{l}\}$ inductively. 

Now we state the main results of \cite{BJMST2}: 
\begin{thm}
The functions $\{h_{n}\}_{n=0}^{\infty}$ defined by 
\eqref{eq:original-solution-1} satisfy the equations 
\eqref{eq:rqkz1}--\eqref{eq:rqkz3} if 
$\omega_{1}(\lambda)$ and $\omega_{2}(\lambda)$ are
solutions of the following system of 
difference equations: 
\begin{eqnarray}
&& 
\omega_{1}(\lambda-1)+\omega_{1}(\lambda)+p_{1}(\lambda)=0, \\ 
&& 
\omega_{2}(\lambda-1)+\omega_{2}(\lambda)+\omega_{1}(\lambda)+
p_{1}(\lambda)+p_{2}(\lambda)=0, 
\label{eq:difference-omega1} \\ 
&&
\omega_{1}(-\lambda)=\omega_{1}(\lambda), \quad 
\omega_{2}(-\lambda)=-\omega_{2}(\lambda),
\label{eq:difference-omega2} 
\end{eqnarray}
where
\begin{eqnarray*}
&& 
p_{1}(\lambda):=\frac{3}{4}\frac{1}{[\lambda][\lambda-1]}-
\frac{1}{4}\frac{[3]}{[\lambda-2][\lambda+1]}, \\ 
 && 
p_{2}(\lambda):=\frac{1}{2}\frac{1}{[\lambda-1][\lambda-2]}-
\frac{1}{4}\frac{[2]}{[\lambda-1][\lambda+1]}. 
\end{eqnarray*}
Moreover, $h_{n}$ gives the correlation function \eqref{eq:defh} 
in the massive regime when 
$\omega_{j}(\lambda) \, (j=1,2)$ are given by %\eqref{eq:rho_trig1}. 
\eqref{eq:omegaXXZ}. 
\end{thm}

\subsection{Another formula}\label{app:a-2}

In this subsection we construct an operator 
\begin{eqnarray*}
\widetilde{\Omega}_{n}(\lambda_{1}, \ldots , \lambda_{n}) \in 
{\rm End}(V^{\otimes 2n})
\end{eqnarray*}
and rewrite the formula \eqref{eq:original-solution-1} 
by using $\widetilde{\Omega}_{n}$ 
(see \eqref{eq:omega-formula}). 
%The final result is given by 
%such that 
%\begin{eqnarray*}
%h_{n}(\lambda_{1}, \ldots , \lambda_{n})=2^{-n}
%e^{\widehat{\Omega}_{n}(\lambda_{1}, \ldots ,\lambda_{n})}\,
%{\bf s}_{n}. 
%\end{eqnarray*}
The procedure of rewriting here is similar to 
that in Section 11 of \cite{BJMST2}, but slightly different. 
%Note that, while we use the same notation as \cite{BJMST2}, 
%the quantities are different. 

First note that the vector $s_{1, \bar{2}}s_{\bar{1}, 2}$ 
in the right hand side of \eqref{eq:def-X} can be 
replaced by $s_{1, 2}s_{\bar{1}, \bar{2}}$ 
because of the following reason. 
We have the equality 
\begin{eqnarray*}
s_{1, \bar{2}}s_{\bar{1}, 2}=
s_{1, 2}s_{\bar{1}, \bar{2}}-s_{1, \bar{1}}s_{2, \bar{2}}. 
\end{eqnarray*}
{}From the crossing symmetry
\begin{eqnarray*}
\ell_{\bar{2}}(\frac{\lambda_{1,2}}{2}-1)\,s_{2, \bar{2}}=
{}-\ell_{2}(-\frac{\lambda_{1,2}}{2})\,s_{2,\bar{2}}  
\end{eqnarray*}
and the cyclicity of the trace function, we see that 
\begin{eqnarray*}
&& 
{\rm Tr}_{\lambda_{1,2}}
(t_{n}^{[1]}(\frac{\lambda_{1}+\lambda_{2}}{2}))\,s_{2,\bar{2}} \\ 
&& {}=-
{\rm Tr}_{\lambda_{1,2}}( 
\ell_{\bar{3}}(\frac{\lambda_{1}+\lambda_{2}}{2}-\lambda_{3}-1) 
\cdots 
\ell_{3}(\frac{\lambda_{1}+\lambda_{2}}{2}-\lambda_{3}) \, 
\ell_{2}(\frac{\lambda_{1,2}}{2}) 
\ell_{2}(-\frac{\lambda_{1,2}}{2})) \, 
s_{2,\bar{2}}.
\end{eqnarray*}
Use the quantum determinant formula 
\begin{eqnarray}
{\rm Tr}_{d}(x \, \ell(\lambda)\ell(-\lambda))=
\left[\frac{d}{2}+\lambda\right]
\left[\frac{d}{2}-\lambda\right] 
{\rm Tr}_{d}(x) \qquad 
(\forall{x} \in U_{q}(sl_{2})).
\label{eq:app-quantumdet}
\end{eqnarray}
Then we find 
\begin{eqnarray*}
{\rm Tr}_{\lambda_{1,2}}
(t_{n}^{[1]}(\frac{\lambda_{1}+\lambda_{2}}{2}))\,s_{2,\bar{2}}=0. 
\end{eqnarray*}
As a result we obtain 
\begin{eqnarray*}
{\rm Tr}_{\lambda_{1,2}}
(t_{n}^{[1]}(\frac{\lambda_{1}+\lambda_{2}}{2}))\,s_{1,\bar{2}}s_{\bar{1},2}=
{\rm Tr}_{\lambda_{1,2}}
(t_{n}^{[1]}(\frac{\lambda_{1}+\lambda_{2}}{2}))\,s_{1,2}s_{\bar{1},\bar{2}}.
\end{eqnarray*}
In the following we use the formula for ${}_{n}X_{n-2}^{(i,j)}$ where 
$s_{1,\bar{2}}s_{\bar{1},2}$ is replaced by 
$s_{1,2}s_{\bar{1},\bar{2}}$.

Introduce the operator 
${}_{n-1}\tilde{\Pi}_{n} \in {\rm End}(V^{\otimes 2n}, V^{\otimes
2(n-1)})$ defined by 
\begin{eqnarray*}
\mathcal{P}_{n, \bar{n}}^{-}\,u=
({}_{n-1}\tilde{\Pi}_{n}u)_{1, \ldots , n-1, \overline{n-1}, \ldots ,
\overline{1}}\,
s_{n, \bar{n}},  
\end{eqnarray*}
and set ${}_{n-2}\tilde{\Pi}_{n}:=
 {}_{n-2}\tilde{\Pi}_{n-1} \cdot {}_{n-1}\tilde{\Pi}_{n}$.

Now we define the operator 
$\widetilde{X}_{n}^{(i,j)} \in {\rm End}(V^{\otimes 2n})$ by
\begin{eqnarray}
&& 
\widetilde{X}_{n}^{(i,j)}(\lambda_{1}, \ldots , \lambda_{n}) 
\label{eq:defXhatpre} \\ 
&& :=-4\,
{}_{n}X_{n-2}^{(i, j)}(\lambda_{1}, \ldots , \lambda_{n}) \cdot 
{}_{n-2}\tilde{\Pi}_{n} \cdot \overrightarrow{P}^{(i,j)}_{n}\, 
\overrightarrow{\mathbb{R}}^{(i, j)}_{n}(\lambda_{1}, \ldots ,
\lambda_{n}),
\nonumber 
\end{eqnarray}
where 
\begin{eqnarray*}
\overrightarrow{P}_{n}^{(i,j)}:=
P_{n-1, n-2} \cdots P_{i+1,i} \cdot 
P_{n,n-1} \cdots P_{j+1,j} \cdot 
P_{\overline{n-2}, \overline{n-1}} \cdots P_{\overline{i}, \overline{i+1}}
\cdot 
P_{\overline{n-1}, \overline{n}} \cdots P_{\overline{j}, \overline{j+1}}.
\end{eqnarray*}
and 
\begin{eqnarray*}
\overrightarrow{\mathbb{R}}_{n}^{(i, j)}(\lambda_{1}, \ldots , \lambda_{n})
&:=&
R_{i,n}(\lambda_{i,n}) \cdots
\widehat{R_{i,j}(\lambda_{i,j})} \cdots  
R_{i,i+1}(\lambda_{i,i+1}) \\ 
&\times& 
R_{j,n}(\lambda_{j,n}) \cdots 
R_{j,j+1}(\lambda_{j,j+1}) \\ 
&\times& 
R_{\overline{n}, \overline{i}}(\lambda_{n,i}) \cdots 
\widehat{R_{\overline{j}, \overline{i}}(\lambda_{j,i})} \cdots 
R_{\overline{i+1}, \overline{i}}(\lambda_{i+1,i}) \\
&\times& 
R_{\overline{n}, \overline{j}}(\lambda_{n,j}) \cdots 
R_{\overline{j+1}, \overline{j}}(\lambda_{j+1,j}).
\end{eqnarray*}
More explicitly, we have 
\begin{eqnarray}
&& 
\widetilde{X}^{(i, j)}_{n}(\lambda_{1}, \ldots , \lambda_{n}):= 
\frac{-4}
{[\lambda_{i,j}]\prod_{p \not= i,j}[\lambda_{i,p}][\lambda_{j,p}]}
\overleftarrow{\mathbb{R}}_{n}^{(i,j)}
(\lambda_{1}, \ldots ,\lambda_{n}) 
\label{eq:defXhat} \\ 
&& \qquad {}\times 
{\rm Tr}_{\lambda_{i,j}}\left( 
\ell_{\bar{j}}\bigl(\frac{\lambda_{i,j}}{2}-1\bigr) \, 
t_{n}^{[i,j]}\bigl(\frac{\lambda_{i}+\lambda_{j}}{2}; \lambda_{1}, \ldots ,
\lambda_{n}\bigr)\,
\ell_{j}\bigl(\frac{\lambda_{i,j}}{2}\bigr)
\right) 
\nonumber \\ 
&& \qquad {}\times 
P_{\bar{i},j}\mathcal{P}_{i, \bar{i}}^{-}\mathcal{P}_{j, \bar{j}}^{-} 
\overrightarrow{\mathbb{R}}_{n}^{(i,j)}(\lambda_{1}, \ldots , \lambda_{n}),
\nonumber
\end{eqnarray}
where 
\begin{eqnarray*}
&& 
t_{n}^{[i,j]}(\lambda;\lambda_{1}, \ldots , \lambda_{n}) \\ 
&& {}:=
\ell_{\bar{1}}(\lambda-\lambda_{1}-1) \cdots 
\widehat{\ell_{\bar{i}}(\lambda-\lambda_{i}-1)} \cdots  
\widehat{\ell_{\bar{j}}(\lambda-\lambda_{j}-1)} \cdots
\ell_{\bar{n}}(\lambda-\lambda_{n}-1)  \\ 
&& {}\times 
\ell_{n}(\lambda-\lambda_{n}) \cdots 
\widehat{\ell_{\bar{j}}(\lambda-\lambda_{j})} \cdots  
\widehat{\ell_{\bar{i}}(\lambda-\lambda_{i})} \cdots
\ell_{1}(\lambda-\lambda_{1}).
\end{eqnarray*}
To obtain \eqref{eq:defXhat} from \eqref{eq:defXhatpre} we used 
\begin{eqnarray*}
\overleftarrow{P}^{(i,j)}_{n}
s_{1,2}s_{\bar{1},\bar{2}}\left( 
{}_{n-2}\tilde{\Pi}_{n} \cdot  
\overrightarrow{P}^{(i,j)}_{n} u 
\right)_{3, \ldots , n, \bar{n}, \ldots , \bar{3}}=
P_{\bar{i},j}\mathcal{P}^{-}_{i, \bar{i}}
\mathcal{P}^{-}_{j, \bar{j}} u
\end{eqnarray*}
for $u \in V^{\otimes 2n}$. 

In the same way as \eqref{eq:defGG} 
the operator $\widetilde{X}_{n}^{(i,j)}$ is decomposed into 
two parts: 
\begin{eqnarray*}
\widetilde{X}_{n}^{(i,j)}(\lambda_{1}, \ldots , \lambda_{n})=
\widetilde{X}_{1, n}^{(i,j)}(\zeta_{1}, \ldots , \zeta_{n})
{}-\lambda_{i,j}\cdot
\widetilde{X}_{2, n}^{(i,j)}(\zeta_{1}, \ldots , \zeta_{n}),
\end{eqnarray*}
where $\widetilde{X}_{a, n}^{(i,j)} \, (a=1,2)$ are
rational functions in $\zeta_{1}, \ldots , \zeta_{n}$ 
which take values in ${\rm End}(V^{\otimes 2n})$. 
Then take solutions $\omega_{1}(\lambda)$ and 
$\omega_{2}(\lambda)$ of the equations 
\eqref{eq:difference-omega1}--\eqref{eq:difference-omega2}, 
and define 
\begin{eqnarray*}
\widetilde{\Omega}^{(i,j)}_{n}(\lambda_{1} , \ldots, \lambda_{n}):=
%{}-4\left( 
\sum_{a=1,2}
\omega_{a}(\lambda_{i,j})
\widetilde{X}^{(i,j)}_{a, n}(\zeta_{1}, \ldots, \zeta_{n}). 
%+\omega(\lambda_{i,j})W_{n}^{(i,j)}(\zeta_{1}, \ldots , \lambda_{n}) 
%\right). 
\end{eqnarray*}
By definition we set 
$\widetilde{\Omega}_{n}^{(j,i)}=\widetilde{\Omega}_{n}^{(i,j)}$ 
for $1 \le i<j \le n$.
Then the operator $\widetilde{\Omega}_{n}^{(i,j)}$ has the following properties:

\begin{prop}\label{prop:Omegarewriting}
(1)\, 
Suppose that $i<j$ and $k<l$. 
If $\{i, j\}\cap\{k,l\}\not=\emptyset $, 
we have 
\begin{eqnarray*}
\widetilde{\Omega}^{(i,j)}_{n}(\lambda_{1},\ldots , \lambda_{n}) \cdot 
{}_{n}\Omega_{n-2}^{(k,l)}(\lambda_{1}, \ldots , \lambda_{n})=0.
\end{eqnarray*}
If $\{i, j\}\cap\{k,l\}=\emptyset $, 
we have 
\begin{eqnarray*}
&& 
\widetilde{\Omega}^{(i,j)}_{n}(\lambda_{1},\ldots , \lambda_{n}) \cdot 
{}_{n}\Omega_{n-2}^{(k,l)}(\lambda_{1}, \ldots , \lambda_{n}) \\ 
&& {}=
{}-4{}_{n}\Omega_{n-2}^{(i,j)}(\lambda_{1}, \ldots , \lambda_{n}) \cdot 
{}_{n-2}\Omega_{n-4}^{(k',l')}
(\lambda_{1}, \ldots , \widehat{\lambda_{i}}, \ldots ,
\widehat{\lambda_{j}}, \ldots , \lambda_{n})  \\ 
&& \quad {}\times 
{}_{n-4}\tilde{\Pi}_{n-2} \cdot 
\overrightarrow{P}^{(i', j')}_{n-2} 
\overrightarrow{\mathbb{R}}^{(i', j')}_{n-2}
(\lambda_{1}, \ldots , \widehat{\lambda_{k}}, \ldots ,
\widehat{\lambda_{l}}, \ldots , \lambda_{n}), 
\end{eqnarray*} 
where $k'$ and $l'$ are the positions of 
$\lambda_{k}$ and $\lambda_{l}$ in 
$(\lambda_{1}, \ldots , \widehat{\lambda_{i}}, \ldots ,
\widehat{\lambda_{j}}, \ldots , \lambda_{n})$, and 
$i'$ and $j'$ are the positions of 
$\lambda_{i}$ and $\lambda_{j}$ in 
$(\lambda_{1}, \ldots , \widehat{\lambda_{k}}, \ldots ,
\widehat{\lambda_{l}}, \ldots , \lambda_{n})$.
\medskip 

\noindent(2)\, 
The operators $\widetilde{\Omega}_{n}^{(i,j)}$ satisfy 
the exchange relations:
\begin{eqnarray*}
&& 
\check{R}_{k, k+1}(\lambda_{k,k+1})
\check{R}_{\overline{k+1}, \overline{k}}(\lambda_{k+1, k}) 
\widetilde{\Omega}_{n}^{(i,j)}
(\ldots , \lambda_{k}, \lambda_{k+1}, \ldots) \\
&& {}=
\widetilde{\Omega}_{n}^{(\pi_{k}(i),\pi_{k}(j))}
(\ldots , \lambda_{k+1}, \lambda_{k}, \ldots)
\check{R}_{k, k+1}(\lambda_{k,k+1})
\check{R}_{\overline{k+1}, \overline{k}}(\lambda_{k+1, k}).  
\end{eqnarray*}
Here $\pi_{k}$ is the transposition $(k, k+1)$.
\medskip 

\noindent(3)\, 
The operators $\widetilde{\Omega}_{n}^{(i,j)}$ are commutative: 
\begin{eqnarray*}
\widetilde{\Omega}_{n}^{(i,j)}\widetilde{\Omega}_{n}^{(k,l)}=
\widetilde{\Omega}_{n}^{(k,l)}\widetilde{\Omega}_{n}^{(i,j)} 
\quad 
\hbox{for all} \quad i<j, \, k<l. 
\end{eqnarray*}

\end{prop}

The proof of Proposition \ref{prop:Omegarewriting} 
is quite similar to that of Lemma 12.1, Lemma 12.2 
and Lemma 12.3 in \cite{BJMST2}. 
In the proof of (1), use the recurrence relation 
\begin{eqnarray}
&&{}_{n-1}\tilde{\Pi}_{n}\cdot 
{}_{n}X_{n-2}^{(i,j)}(\lambda_{1}, \ldots , \lambda_{n})
\label{eq:contraction}\\
&&\quad=
\left\{ 
\begin{array}{ll}
 0 & \hbox{if} \quad j=n, \\
 {}_{n-1}X_{n-3}^{(i,j)}(\lambda_{1}, \ldots ,\lambda_{n-1}) \cdot 
 {}_{n-3}\tilde{\Pi}_{n-2} & \hbox{if} \quad j<n.
\end{array}
\right.
\nonumber
\end{eqnarray}

Now we define the operator $\widetilde{\Omega}_{n}$ by 
\begin{eqnarray*}
\widetilde{\Omega}_{n}(\lambda_{1}, \ldots , \lambda_{n}):=
\sum_{1 \le i<j \le n}
\widetilde{\Omega}_{n}^{(i,j)}
(\lambda_{1}, \ldots , \lambda_{n}). 
\end{eqnarray*}
Then from Proposition \ref{prop:Omegarewriting} and 
the equality 
\begin{eqnarray*}
P_{1,2}P_{\bar{2}, \bar{1}} 
R_{1, 2}(\lambda)R_{\bar{2}, \bar{1}}(-\lambda)
s_{1, \bar{1}}s_{2, \bar{2}}= 
s_{1, \bar{1}}s_{2, \bar{2}}, 
\end{eqnarray*}
we get 
\begin{thm}\label{thm:app-a-1}
The formula \eqref{eq:original-solution-1} for $h_{n}$ 
can be rewritten as follows: %using the operator $\widetilde{\Omega}_{n}$: 
\begin{eqnarray}
h_{n}(\lambda_{1}, \ldots , \lambda_{n})=2^{-n}
e^{\widetilde{\Omega}_{n}(\lambda_{1}, \ldots , \lambda_{n})}
{\bf s}_{n}. 
\label{eq:omega-formula} 
\end{eqnarray} 
\end{thm} 

%\subsection{Rewriting the Ansatz}\label{app:new}

\subsection{Derivation of the new formula}\label{app:new}

Finally we prove that the operator $\Omega_{n}$ defined in \eqref{eq:newformula} 
is equal to $\widetilde{\Omega}_{n}$. 

{}From the definition \eqref{eq:GGt}
of $X_{a, n} \, (a=1, 2)$, we can easily see that 
\begin{eqnarray*}
{\rm res}_{\mu_{1}=\lambda_{j}} 
{\rm res}_{\mu_{2}=\lambda_{j}}
X_{a, n}(\mu_{1}, \mu_{2} ; \lambda_{1}, \ldots , \lambda_{n})=0 
\end{eqnarray*}
and 
\begin{eqnarray*}
&& 
{\rm res}_{\mu_{1}=\lambda_{i}} 
{\rm res}_{\mu_{2}=\lambda_{j}}
X_{a, n}(\mu_{1}, \mu_{2} ; \lambda_{1}, \ldots , \lambda_{n}) \\ 
&& {}=(-1)^{a}
{\rm res}_{\mu_{1}=\lambda_{j}} 
{\rm res}_{\mu_{2}=\lambda_{i}}
X_{a, n}(\mu_{1}, \mu_{2} ; \lambda_{1}, \ldots , \lambda_{n}).
\end{eqnarray*}
The operator 
\begin{eqnarray*}
{\rm Tr}_{2,2}\left( 
T_{n}(\mu_{1}) \otimes T_{n}(\mu_{2}) \cdot B(\mu_{1,2}) 
\right) 
\end{eqnarray*}
is skew-symmetric with respect to $\mu_{1}$ and $\mu_{2}$ 
because of the commutation relation
\begin{eqnarray*}
R_{12}(\mu)B_{12}(\mu)=-B_{21}(-\mu)R_{12}(\mu). 
\end{eqnarray*} 
The functions $\omega_{a} \, (a=1,2)$ satisfy the parity conditions 
\begin{eqnarray*}
\omega_{a}(-\lambda)=(-1)^{a-1}\omega_{a}(\lambda).
\end{eqnarray*} 
By using the above properties we have 
\begin{eqnarray*}
\Omega_{n}(\lambda_{1}, \ldots , \lambda_{n})=
\sum_{1 \le i<j \le n} 
\Omega_{n}^{(i,j)}(\lambda_{1}, \ldots , \lambda_{n}), 
\end{eqnarray*}
where 
\begin{eqnarray*}
\Omega_{n}^{(i,j)}(\lambda_{1}, \ldots , \lambda_{n})&:=& 
\frac{(-1)^{n}}{\kappa^{2}} 
\sum_{a=1, 2}\omega_{a}(\lambda_{i,j})
{\rm res}_{\mu_{1}=\lambda_{i}} 
{\rm res}_{\mu_{2}=\lambda_{j}}
X_{a, n}(\mu_{1}, \mu_{2} ; \lambda_{1}, \ldots , \lambda_{n}) \\ 
&& \hspace{5em} {}\times 
{\rm Tr}_{2,2}(T_{n}(\lambda_{i})\otimes T_{n}(\lambda_{j}) \cdot B(\lambda_{i,j})).
 \end{eqnarray*}
In the following we show the equality  
$\Omega_{n}^{(i,j)}=\widetilde{\Omega}_{n}^{(i,j)}$. 
To this end we prove two lemmas.

Let $B$ be a linear operator acting on 
$\mathbb{C}^{2} \otimes \mathbb{C}^{2}$.
Then 
\begin{eqnarray*}
{\rm Tr}_{2,2}(T_{n}(\lambda_{i}) \otimes T_{n}(\lambda_{j}) \cdot B) 
\end{eqnarray*} 
is the operator acting on 
$V_{1} \otimes \cdots \otimes V_{n} \otimes V_{\bar{n}} \otimes \cdots
\otimes V_{\bar{1}}$.

\begin{lem}\label{lem:app-main1}
We have 
\begin{eqnarray}
&& 
{\rm Tr}_{2,2}(T_{n}(\lambda_{i}) \otimes T_{n}(\lambda_{j}) \cdot B) 
\label{eq:app-lem1} \\
&& {}=4
\overleftarrow{\mathbb{R}}_{n}^{(i,j)}(\lambda_{1}, \ldots , \lambda_{n})
B_{i,j}
\mathcal{P}_{i, \bar{i}}^{-}
\mathcal{P}_{j, \bar{j}}^{-}
\overrightarrow{\mathbb{R}}_{n}^{(i,j)}(\lambda_{1}, \ldots , \lambda_{n}).
\nonumber
\end{eqnarray} 
\end{lem}

\begin{proof}
We prepare some notation. 
Denote by $V_a \otimes V_b \, (V_{a} \simeq V_{b} \simeq \mathbb{C}^{2})$ 
the tensor product of the two-dimensional spaces 
on which the trace ${\rm Tr}_{2,2}$ is taken. 
%and $V_i\simeq V_{\bar  i}\simeq\C^2$ ($1\leq i\leq n$).
Set
\be
V^Q=V_1\otimes\cdots\otimes V_n\otimes
V_{\bar  n}\otimes\cdots\otimes V_{\bar 1}.
\en
Let $R_{a,i}(\lambda)$ be the $R$ matrix acting on the
$a$-th and the $i$-th component of $V_a\otimes V_b\otimes V^Q$.
We set
\be
T_a(\lambda)=R_{a,\bar 1}(\lambda-\lambda_1-1)\cdots
R_{a,\bar  n}(\lambda-\lambda_n-1)
R_{a,n}(\lambda-\lambda_n)\cdots R_{a,1}(\lambda-\lambda_1).
\en
We define $T_b(\lambda)$ similarly. 
%Let $B \in\End(\C^2\otimes\C^2)$
%be a $4\times4$ matrix. depending on $\lambda$.
We denote by $B_{a,b}$ the operator $B$ acting on
$V_a\otimes V_b$. %\simeq\C^2\otimes\C^2$.
Similarly, $B_{i,b}$, %(\lambda)$, 
etc., are defined.

Denote by $X$ the left hand side of \eqref{eq:app-lem1}. 
Then we have 
%Fix $i,j$ such that $1\leq i<j\leq n$.
%Let us calculate
\be
X=\tr_{a,b}(B_{a,b}T_a(\lambda_i)T_b(\lambda_j)).
\en
Here $\tr_{a,b}$ means taking trace on $V_a\otimes V_b$.
We use the following properties of the $R$ matrix.
\bel
&&R_{1,2}(0)=P_{1,2},\\
&&R_{1,2}(-1)=-2\P^-_{1,2},\\
&&R_{1,2}(\lambda_{1,2})R_{1,3}(\lambda_{1,3})R_{2,3}(\lambda_{2,3})
=R_{2,3}(\lambda_{2,3})R_{1,3}(\lambda_{1,3})R_{1,2}(\lambda_{1,2}),
\label{YB}\\
&&R_{1,2}(\lambda)R_{2,1}(-\lambda)=1,\label{UNI}\\
&&R_{1,3}(\lambda)R_{1,2}(-1)=
{}-R_{3,2}(-1-\lambda)R_{1,2}(-1),\label{CR1}\\
&&R_{1,2}(-1)R_{1,3}(\lambda)=
{}-R_{1,2}(-1)R_{3,2}(-1-\lambda).\label{CR2}
\enl

We abbreviate the arguments of the $R$-matrices.
They can be understood from the space indices $a,b,1,\ldots,n,\bar 1,\ldots,\bar  n$
through the correspondences
$a\leftrightarrow\lambda_i$,
$b\leftrightarrow\lambda_j$,
$1\leftrightarrow\lambda_1$,\ldots,$n\leftrightarrow\lambda_n$,
$\bar 1\leftrightarrow\lambda_1+1$,\ldots,$\bar  n\leftrightarrow\lambda_n+1$.
For example, by $R_{i,\bar  j}$ we mean $R_{i,\bar  j}(\lambda_{i,j}-1)$.
We extend this convention to $T_a$ and $T_b$.
We also use the abbreviation
\be
R_{i,[i-1,1]}=R_{i,i-1}R_{i,i-2}\cdots R_{i,1},
\en
etc..

With the above convention, we have
\be
&&X=
\tr_{a,b}\{B_{a,b}\\
&&\times
R_{a,[\bar 1,\bar {i-1}]}
R_{a,\bar  i}(-1)
R_{a,[\bar {i+1},\bar {j-1}]}
R_{a,\bar  j}
R_{a,[\bar {j+1},\bar  n]}
R_{a,[n,j+1]}
R_{a,j}
R_{a,[j-1,i+1]}
P_{a,i}
R_{a,[i-1,1]}\\
&&\times\,
R_{b,[\bar 1,\bar {i-1}]}
R_{b,\bar  i}\,
R_{b,[\bar {i+1},\bar {j-1}]}
R_{b,\bar  j}(-1)\,
R_{b,[\bar {j+1},\bar  n]}
R_{b,[n,j+1]}
P_{b,j}\,
R_{b,[j-1,i+1]}
R_{b,i}
R_{b,[i-1,1]}\}.
\en
We write the argument $-1$ in two places in order to emphasize
where we can use the crossing symmetries (\ref{CR1}), (\ref{CR2}).
Using the cyclicity of the trace, we bring 
$R_{a,[i-1,1]}R_{b,[i-1,1]}$ to the left of 
$B_{a,b}$. Then, we move $P_{a,i}$ to the left
while we change all $a$ to $i$. Finally, we can
eliminate $\tr_aP_{a,i}$ because this is equal to the identity operator.
Thus, we get
\be
&&X=
R_{i,[i-1,1]}\tr_b\{
R_{b,[i-1,1]}B_{i,b}\\
&&\times
R_{i,[\bar 1,\bar {i-1}]}
R_{i,\bar  i}(-1)
R_{i,[\bar {i+1},\bar {j-1}]}
R_{i,\bar  j}
R_{i,[\bar {j+1},\bar  n]}
R_{i,[n,j+1]}
\underline{R_{i,j}R_{i,[j-1,i+1]}}\\
&&\times\,
R_{b,[\bar 1,\bar {i-1}]}
R_{b,\bar  i}\,
R_{b,[\bar {i+1},\bar {j-1}]}
R_{b,\bar  j}(-1)\,
R_{b,[\bar {j+1},\bar  n]}
R_{b,[n,j+1]}
\underline{P_{b,j}}\,
R_{b,[j-1,i+1]}
R_{b,i}\}
\en
Using the crossing symmetries (\ref{CR1}), (\ref{CR2}), and pushing 
the underlined terms to the right, we obtain
\be
&&X=
R_{i,[i-1,1]}
R_{[\bar {i-1},\bar 1],\bar i}\tr_b\{
R_{b,[i-1,1]}B_{i,b}\\
&&\times
R_{i,\bar  i}(-1)
R_{[\bar  n,\bar {j+1}],\bar  i}
R_{\bar  j,\bar  i}
R_{[\bar {j-1},\bar  {i+1}],\bar  i}
R_{i,[n,j+1]}
\\
&&\times\,
R_{[\bar {j-1},\bar {i+1}],\bar  j}
R_{\bar  i,\bar  j}\,
R_{[\bar {i-1},\bar 1],\bar  j}
R_{b,\bar  j}(-1)\,
R_{[\bar  n,\bar {j+1}],\bar  j}
R_{b,[n,j+1]}
\underline{R_{i,j}R_{i,[j-1,i+1]}}
R_{j,[j-1,i+1]}
R_{j,i}
\underline{P_{b,j}}\,\}
\en
Using the Yang-Baxter equation (\ref{YB}) and
the unitarity relation (\ref{UNI}), we reduce $R_{ij}R_{ji}$,
and $R_{\bar  j,\bar  i}R_{\bar  i,\bar  j}$.
\be
&&%[1-\lambda_{i,j}]^{-2}[1+\lambda_{i,j}]^{-2}
X=
R_{i,[i-1,1]}
R_{[\bar {i-1},\bar 1],\bar i}\tr_b\{
R_{b,[i-1,1]}B_{i,b}\\
&&\times
R_{i,\bar  i}(-1)
R_{[\bar  n,\bar {j+1}],\bar  i}
R_{i,[n,j+1]}
\underline{R_{[\bar {j-1},\bar {i+1}],\bar  j}}
R_{[\bar {j-1},\bar  {i+1}],\bar  i}\\
&&\times
R_{[\bar {i-1},\bar 1],\bar  j}
R_{b,\bar  j}(-1)\,
R_{[\bar  n,\bar {j+1}],\bar  j}
R_{b,[n,j+1]}
\underline{R_{j,[j-1,i+1]}}
R_{i,[j-1,i+1]}
P_{b,j}\}
\en
The underlined terms can be brought outside the trace.
Therefore, handling $P_{b,j}$ in the same way as $P_{a,i}$, we obtain
\be
&&%[1-\lambda_{i,j}]^{-2}[1+\lambda_{i,j}]^{-2}
X=
R_{i,[i-1,1]}
R_{[\bar {i-1},\bar 1],i}
\underline{R_{[\bar {j-1},\bar {i+1}],\bar  j}R_{j,[j-1,i+1]}}
R_{j,[i-1,1]}B_{i,j}\\
&&\times
R_{i,\bar  i}(-1)
R_{[\bar  n,\bar {j+1}],\bar  i}
R_{i,[n,j+1]}
R_{[\bar {j-1},\bar  {i+1}],\bar  i}\\
&&\times
R_{[\bar {i-1},\bar 1],\bar  j}
R_{j,\bar  j}(-1)\,
R_{[\bar  n,\bar {j+1}],\bar  j}
R_{j,[n,j+1]}
R_{i,[j-1,i+1]}\\
&&=
%\buildrel\leftarrow\over R_{i,j}
\overleftarrow{\mathbb{R}}_{n}^{(i, j)}
B_{i,j}R_{i,\bar  i}(-1)R_{j,\bar  j}(-1)
%\buildrel\rightarrow\over R_{i,j}
\overrightarrow{\mathbb{R}}^{(i,j)}_{n}.
\en
This completes the proof.
\end{proof}

To state the second lemma we introduce some notation. 
The $\ell$-operator is written as 
\begin{eqnarray*}
\ell(\lambda)=\sum_{\alpha=0}^{3}
\ell^{\alpha}(\lambda) \otimes \sigma^{\alpha} 
\end{eqnarray*}
where 
\be
\ell^{0}(\lambda)&=&\frac{\bigl(q^{\lambda+\frac12}-q^{-\lambda-\frac12}\bigr)
\bigl(q^{\frac H2}+q^{-\frac H2}\bigr)}{2(q-q^{-1})},\\
\ell^{1}(\lambda)&=&\frac{q^{\frac{1-H}2}E+Fq^{\frac{(H-1)}2}}2,\\
i\ell^{2}(\lambda)&=&\frac{q^{\frac{1-H}2}E-Fq^{\frac{(H-1)}2}}2,\\
\ell^{3}(\lambda)&=&\frac{\bigl(q^{\lambda+\frac12}+q^{-\lambda-\frac12}\bigr)
\bigl(q^{\frac H2}-q^{-\frac H2}\bigr)}{2(q-q^{-1})}.
\en
For $\alpha, \beta=0,1,2,3$ we set 
\begin{eqnarray*}
\epsilon_{\alpha\beta}=\left\{ 
\begin{array}{cl}
i & \hbox{if} \,\, (\alpha, \beta)=(1,2), (2,3), (3,1), \\
{}-i & \hbox{if} \,\, (\alpha, \beta)=(2,1), (3,2), (1,3), \\
1 & \hbox{otherwise.}
\end{array}
\right.
\end{eqnarray*}

\begin{lem}\label{lem:app-main2}
Let $B \in {\rm End}(\mathbb{C}^{2} \otimes \mathbb{C}^{2})$ 
be an operator %given in the following form:
\begin{eqnarray*}
B=\sum_{\alpha, \beta=0}^{3}\epsilon_{\alpha\beta}B^{\alpha\beta} 
(\sigma^{\alpha} \otimes \sigma^{\beta}) \qquad 
(B^{\alpha\beta} \in \mathbb{C}). 
\end{eqnarray*}
Then we have 
\begin{eqnarray}
&& 
\hspace{3em}
{\rm Tr}_{\lambda_{i,j}}\left(
t_{n}(\frac{\lambda_{i}+\lambda_{j}}{2})\right)
\overleftarrow{\mathbb{R}}_{n}^{(i,j)}
(\lambda_{1}, \ldots , \lambda_{n}) 
B_{i, j} 
\mathcal{P}_{i, \bar{i}}^{-} \mathcal{P}_{j, \bar{j}}^{-} 
\label{eq:app-lem2} \\ 
&& {}=
\overleftarrow{\mathbb{R}}_{n}^{(i,j)}
(\lambda_{1}, \ldots , \lambda_{n}) 
{\rm Tr}_{\lambda_{i,j}}\left(
\ell_{\bar{j}}(\frac{\lambda_{i,j}}{2}-1) 
t_{n}^{[i,j]}(\frac{\lambda_{i}+\lambda_{j}}{2})
\ell_{j}(\frac{\lambda_{i,j}}{2}) \cdot Y(\lambda_{i,j})
\right)
P_{\bar{i}, j} \mathcal{P}_{i, \bar{i}}^{-} \mathcal{P}_{j, \bar{j}}^{-},
\nonumber
\end{eqnarray} 
where $Y(\lambda)$ is given by 
\begin{eqnarray}
Y(\lambda):=-4\sum_{\alpha,\beta=0}^{3}
\epsilon_{\alpha\beta} C^{\alpha\beta} \, \ell^{\beta}(\frac{\lambda}{2}-1) 
\, \ell^{\alpha}(-\frac{\lambda}{2}-1). 
\label{eq:def-Y}
\end{eqnarray} 
Here the coefficients $C^{\alpha\beta}$ are determined from 
the equality  
\begin{eqnarray}
\begin{pmatrix}
C^{00}\\C^{11}\\C^{22}\\C^{33}\\
\end{pmatrix}
=
\frac12\begin{pmatrix}
1&-1&-1&-1\\
1&-1&\phantom{-}1&\phantom{-}1\\
1&\phantom{-}1&-1&\phantom{-}1\\
1&\phantom{-}1&\phantom{-}1&-1\\
\end{pmatrix}
\begin{pmatrix}
B^{00}\\B^{11}\\B^{22}\\B^{33}\\
\end{pmatrix}
\label{eq:app-44mat}
\end{eqnarray}
and ones obtained by 
changing the indices $(00,11,22,33)$ in the both hand sides 
of \eqref{eq:app-44mat} to 
$(01,10,23,32), (02,13,20,31)$ and $(03,12,21,30)$.
In particular, if $B$ is the identity, the left hand side of 
\eqref{eq:app-lem2} becomes zero. 
\end{lem}

\begin{proof}
We start from the left hand side of \eqref{eq:app-lem2}.
The $R$-matrices $\overleftarrow{\mathbb{R}}^{(i,j)}_{n}$ go through the
$\ell$-operators $t_n$:
\bea
&&\overleftarrow{\mathbb{R}}^{(i,j)}_{n}
{\rm Tr}_{\lambda_{i,j}}\Bigl(
\ell_{\bar  i}\left(\scriptstyle\frac{\lambda_{j,i}}2-1\right)
\ell_{\bar  j}\left(\scriptstyle\frac{\lambda_{i,j}}2-1\right)
t_{n}^{[i,j]}({\scriptstyle\frac{\lambda_i+\lambda_j}{2}})
\ell_j\left(\scriptstyle\frac{\lambda_{i,j}}2\right)
\ell_i\left(\scriptstyle\frac{\lambda_{j,i}}2\right)
B_{i,j}\Bigr)\P^-_{i,\bar i}\P^-_{j,\bar j}\,.
\label{eq:ambiguity1}
\ena
Now it is easy to see that the above operator becomes zero 
if $B$ is the identity because of 
the cyclicity of the trace, the crossing symmetry 
\begin{eqnarray*}
\ell_1(\lambda)\P^-_{1,2}&=&-\ell_2(-\lambda-1)\P^-_{1,2}, %\label{CR3}\\ 
\end{eqnarray*}
and 
the quantum determinant formula \eqref{eq:app-quantumdet}. 

%Therefore, 
Let us proceed the calculation of \eqref{eq:ambiguity1}.
Without loss of generality, we consider the case $i=1,j=2$:
\be
Z:={\rm Tr}_{\lambda_{1,2}}\Bigl(
\ell_{\bar1}\left(\scriptstyle\frac{\lambda_{2,1}}2-1\right)
\ell_{\bar2}\left(\scriptstyle\frac{\lambda_{1,2}}2-1\right)
t^{[1,2]}_x\left(\lambda\right)
\ell_2\left(\scriptstyle\frac{\lambda_{1,2}}2\right)
\ell_1\left(\scriptstyle\frac{\lambda_{2,1}}2\right)
\Bigr)B_{1,2}\P^-_{1,\bar1}\P^-_{2,\bar2}\,,
\en
where
\be
&&\lambda=\frac{\lambda_1+\lambda_2}2.
\en

We rewrite the last part of $Z$ by using 
\be
B_{1,2}\P^-_{1,\bar1}\P^-_{2,\bar2}
=\sum_{\alpha, \beta=0}^3 \epsilon_{\alpha\beta} 
C^{\alpha\beta}\sigma^\alpha_{\bar1}\sigma^\alpha_2P_{\bar1,2}\P^-_{1,\bar1}\P^-_{2,\bar2}.
\en
Note that
\be
P_{\bar{1},2}\P^-_{1,\bar1}\P^-_{2,\bar2}=\P^-_{1,2}\P^-_{\bar1,\bar2}P_{\bar1,2}.
\en
Using the crossing symmetry, %(\ref{CR3}), 
we rewrite $Z$ as
\begin{eqnarray*}
&& 
Z=\sum_{\alpha, \beta=0}^3 \epsilon_{\alpha\beta}
C^{\alpha\beta} \\ 
&& {}\times 
{\rm Tr}_{\lambda_{1,2}}\Bigl(
\ell_{\bar1}\left(\scriptstyle\frac{\lambda_{2,1}}2-1\right)
\sigma^\alpha_{\bar1}
\ell_{\bar1}\left(\scriptstyle\frac{\lambda_{2,1}}2\right)
\P^-_{\bar1,\bar2}\cdot
t^{[1,2]}_n\left(\lambda\right)\cdot
\ell_2\left(\scriptstyle\frac{\lambda_{1,2}}2\right)
\sigma^\beta_2\ell_2\left(\scriptstyle\frac{\lambda_{1,2}}2-1\right)
\P^-_{1,2}\Bigr)P_{\bar1,2}. 
\end{eqnarray*}
Then \eqref{eq:app-lem2} follows from 
the quantum determinant formula \eqref{eq:app-quantumdet}, 
the following Proposition \ref{prop:app-mm}, and  
the equality 
\begin{eqnarray*}
\sigma^2\ell(\lambda)^t\cdot\sigma^2&=&-\ell(-\lambda-1). %,\label{CR4}
\end{eqnarray*}
\end{proof}

\begin{prop}\label{prop:app-mm}
Suppose that $m=\sum_{\alpha=0}^3m_\alpha\sigma^\alpha$ where $m_\alpha$ is a scalar
with respect to the $2$-dimensional space on which the Pauli matrices act.
We have
\be
\sigma^\alpha m=-\varepsilon_\alpha(\sigma^2m^t\cdot\sigma^2)\sigma^\alpha+2m_\alpha,\\
m\sigma^\alpha=-\varepsilon_\alpha\sigma^\alpha(\sigma^2m^t\cdot\sigma^2)+2m_\alpha,
\en
where 
$m^{t}=\sum_{\alpha=0}^3m_\alpha(\sigma^\alpha)^{t}$ and 
\begin{eqnarray*}
\varepsilon_\alpha=\begin{cases}1&\hbox{if }\alpha=0;\\-1&\hbox{otherwise}.\end{cases} 
\end{eqnarray*}
\end{prop}

The proof of Proposition \ref{prop:app-mm} is straightforward. 
\medskip 

Now let us prove that 
$\Omega_{n}^{(i,j)}=\widetilde{\Omega}_{n}^{(i,j)}$. 
If $B$ is the operator $B(\mu)$ defined in \eqref{eq:Aab}, 
the corresponding operator $Y(\lambda)$ determined by \eqref{eq:def-Y}
becomes the identity. 
Therefore, from the formula \eqref{eq:defXhat}, we obtain 
\begin{eqnarray*}
&&
\frac{(-1)^{n}}{\kappa^{2}}
{\rm res}_{\mu_{1}=\lambda_{i}} {\rm res}_{\mu_{2}=\lambda_{j}}
{\rm Tr}_{\mu_{1,2}}( T_{n}(\frac{\mu_{1}+\mu_{2}}{2})) \cdot
{\rm Tr}_{2,2}(T_{n}(\lambda_{i}) \otimes T_{n}(\lambda_{j}) \cdot B(\lambda_{i,j})) \\
%\label{eq:residue-omega-partial} \\ 
&& {}=\tilde{X}^{(i,j)}_{n}(\lambda_{1}, \ldots , \lambda_{n}).
%\frac{-4}{[\lambda_{i,j}]^{2}\prod_{p \not= i,j}[\lambda_{i,p}][\lambda_{j,p}]}
%\overleftarrow{\mathbb{R}}_{n}^{(i,j)}
%(\lambda_{1}, \ldots , \lambda_{n}) \\
%&& {}\times 
%{\rm Tr}_{\lambda_{i,j}}\left( 
%\ell_{\bar{j}}(\frac{\lambda_{i,j}}{2}-1) 
%t_{n}^{[i,j]}(\frac{\lambda_{i}+\lambda_{j}}{2}) 
%\ell_{j}(\frac{\lambda_{i,j}}{2}) \right) 
%P_{\bar{i}, j}\mathcal{P}_{i, \bar{i}}^{-} 
%\mathcal{P}_{j, \bar{j}}^{-} 
%\overrightarrow{\mathbb{R}}_{n}^{(i,j)}
%(\lambda_{1}, \ldots , \lambda_{n}).
\end{eqnarray*}
This implies the equality $\Omega_{n}^{(i,j)}=\widetilde{\Omega}_{n}^{(i,j)}$.

The proof of \eqref{eq:GGt2}, \eqref{eq:Bsigma} is similar.
It is easy to see that for the operator $B^{q}(\mu)$ 
%written in the form \eqref{eq:operator-weightzero}, 
%and then we have 
we have $Y(\lambda)=q^{H}$.
Therefore, the right hand side of \eqref{eq:newformula} 
where $X_{a,n}(\mu_{1}, \mu_{2})$ and 
$B(\mu)$ are replaced by $X_{a,n}^{q}(\mu_{1}, \mu_{2})$ and $B^{q}(\mu)$, respectively, 
is equal to $\Omega_{n}$.

Finally we give a sketch of the calculation in the elliptic case. 
The $\ell$-operator is given by 
\begin{eqnarray*}
\ell(\lambda)&=&\sum_{\alpha=0}^3w_a(\lambda)S_\alpha\otimes\sigma^\alpha. 
\end{eqnarray*}
Here $S_\alpha$ ($\alpha=0,1,2,3$) are the generators of
%Sklyanin's algebra and 
Sklyanin algebra and 
\be
w_\alpha(\lambda)=\frac{\theta_{\alpha+1}(2t+\eta)}{2\theta_{\alpha+1}(\eta)},
\en
where $t=\lambda\eta$. 
If we set 
\be
B(\lambda)
=-\frac{\theta_1(2t)\theta_1(4\eta)}
{4\theta_1(2t+2\eta)\theta_1(2t-2\eta)}
\sum_{\alpha=1}^3
\frac{\theta_{\alpha+1}(2t)}{\theta_{\alpha+1}(2\eta)}
\sigma^\alpha \otimes \sigma^\alpha, 
\en
the corresponding operator $Y(\lambda)$ becomes 
%some scalar multiple of the Casimir element
\begin{eqnarray*}
Y(\lambda)&=&
\frac{1}{4}\Bigl\{
\frac{\theta_{1}(2t)}{\theta_{1}(2t-2\eta)}\left( 
\frac{\theta_{1}(t-3\eta)\theta_{1}(t-\eta)}{\theta_{1}^{2}(\eta)}K_{0}-
\frac{\theta_{1}^{2}(t-2\eta)}{\theta_{1}^{2}(\eta)}K_{2}
\right) \\ 
&-&
\frac{\theta_{1}(4\eta)}{2\theta_{1}(2t-2\eta)\theta_{1}(2t-2\eta)\theta_{1}(2\eta)}
\left( 
\frac{\theta_{1}(t-\eta)\theta_{1}(t+\eta)}{\theta_{1}^{2}(\eta)}K_{0}-
\frac{\theta_{1}^{2}(t)}{\theta_{1}^{2}(\eta)}K_{2}
\right)
\Bigr\}, 
\end{eqnarray*}
where $K_{0}$ and $K_{2}$ are the Casimir elements
\eqref{eq:Casimir}.
%\be
%K_{0}=\sum_{\alpha=0}^{3}S_{\alpha}^{2}, \qquad 
%K_2=\sum_{a=1}^3
%\frac{\theta_{a+1}(2\eta)\theta_{a+1}(0)}{\theta_{a+1}^{2}(\eta)}S_{a}^2.
%\en
%Under the trace functional ${\rm Tr}_{d}$ 
%%in which the dimension parameter is equal to $d$, 
%we have
%\be
%K_{0}=\frac{4\theta_1^{2}((d+1)\eta)}{\theta_1^2(2\eta)}, \qquad 
%K_2=\frac{4\theta_1((d+1)\eta)\theta_1((d-1)\eta)}{\theta_1^2(2\eta)}.
%\en
Therefore we obtain 
\begin{eqnarray*}
{\rm Tr}_{\lambda}(x \, Y(\lambda))=\frac{\theta_{1}(2t)}{\theta_{1}(2\eta)}
{\rm Tr}_{\lambda}(x) 
\end{eqnarray*}
for any element $x$ of the Sklyanin algebra. 
This gives the formula \eqref{eq:Omega_ell}.

\section{Relation with the vertex operator approach}
\label{app:book}

Correlation functions of the XXZ model in the massive regime 
have been studied in the framework of representation theory
\cite{JMMN}, \cite{JM}.   
The description in the present paper 
differs from the above literature by a few minor points. 
In this appendix, we compare the two in some detail. 

Consider an inhomogeneous six-vertex model where 
an inhomogeneity parameter $\zeta_j$ is attached to each column $j$.  
By correlation functions we mean those of local 
operators on a single row. 
There are two equivalent formulations  
depending on whether one uses row-to-row or column-to-column transfer matrices. 
In this paper, correlation functions are expressed as 
expectation values with respect to the ground state 
of row-to-row transfer matrices. The latter, and hence the ground state
vector, depend on the $\zeta_j$ while local operators do not.  
In \cite{JM}, on the other hand, 
column-to-column transfer matrices are employed. 
Their ground state vectors are independent of $\zeta_j$. 
The inhomogeneity is encoded rather in local operators, 
expressed in the form of an insertion of half-column transfer matrices. 

Another minor difference between \cite{JM} and the present paper 
is that the $R$ matrices and Hamiltonians are not identical. 
The parameter $q=e^{\pi i \nu}$ of the present paper 
and the corresponding parameter $q_{JM}$ in \cite{JM} are related by 
\be
q_{JM}=-q. 
\en
With this identification, the $R$ matrix $R_{JM}(\zeta)$ 
and the Hamiltonian $H_{JM}$ in \cite{JM} 
are related to $R(\lambda)$ \eqref{eq:def-R} and 
$H_{XXZ}$ \eqref{eq:XXZ} by the gauge transformation 
\bel
R(\lambda)=
(\sigma^3\otimes1)R_{JM}(\zeta)(1\otimes \sigma^3),
\quad 
H_{XXZ}=KH_{JM}K^{-1}, 
\label{RGAUGE}
\enl
where $K=\prod_{\scriptstyle {j:even}}\sigma_j^3$. 

Correlation functions in the present paper are 
related to the mean value of the 
two expectation values with respect to the two 
vectors $|\rm vac\rangle_{(i)}$ considered in \cite{JM}. 
Taking into account the gauge transformation (\ref{RGAUGE}), we have
\bel
&&\prod_{j=1}^n(-\eb_{j})\langle\hbox{vac}|(E_{-\eb_{1},\e_1})_1\cdots
(E_{-\eb_{n},\e_n})_n|\hbox{vac}\rangle\label{PB}\\
&&=\frac12\sum_{i=0,1}\prod_{j=1}^n(-\eb_{j})
\cdot 
{}_{(i)}\langle\hbox{vac}|
%({\scriptstyle \prod_{j\hbox{\tiny : even}}}\s^z_j)
K\cdot (E_{-\eb_{1},\e_1})_1\cdots
(E_{-\eb_{n},\e_n})_n
%({\scriptstyle \prod_{j\hbox{\tiny : even}}}\s^z_j)
\cdot K^{-1} |\hbox{vac}\rangle_{(i)}
\nonumber\\
&&=\frac12\sum_{i=0,1}
\prod_{j\hbox{\tiny : even}}\e_j\prod_{j\hbox{\tiny : odd}}(-\eb_{j})
{}_{(i)}\langle\hbox{vac}|
(E_{-\eb_{1},\e_1})_1\cdots(E_{-\eb_{n},\e_n})_n
|\hbox{vac}\rangle_{(i)}.\nonumber
\enl

The correlation functions 
\begin{eqnarray*}
{}_{(i)}\bra{{\rm vac}}
(E_{\epsilon_{1}, \overline{\epsilon}_{1}})_{1} 
\cdots 
(E_{\epsilon_{n}, \overline{\epsilon}_{n}})_{n} 
\ket{{\rm vac}}_{(i)}
\end{eqnarray*}
can be constructed 
in terms of the vertex operators arising from 
representation theory of the quantum affine algebra 
$U_{-q}=U_{-q}(\slth)$ as follows (recall that $q_{JM}=-q$).
 
Denote by $\Lambda_{i} \, (i=0, 1)$ the fundamental weights 
of $U_{-q}$. 
Let $V(\Lambda_{i})$ be the irreducible highest weight module 
with highest weight $\Lambda_{i}$, 
and $V_{\zeta}=V \otimes \mathbb{C}[\zeta, \zeta^{-1}]$ 
the evaluation module in the principal picture. 
The vertex operator (of type I) is an intertwiner 
\begin{eqnarray*}
\Phi(\zeta) \, : \, 
V(\Lambda_{i}) \longrightarrow V(\Lambda_{1-i}) \otimes V_{\zeta}.
\end{eqnarray*}
Define the components $\Phi_{\epsilon}(\zeta) \, (\epsilon=\pm)$ by 
\begin{eqnarray*}
\Phi_{\epsilon}(\zeta)\,:\, V(\Lambda_{i}) \longrightarrow 
 V(\Lambda_{1-i}), \quad  
\Phi(\zeta)\,u=\sum_{\epsilon=\pm}
\left(\Phi_{\epsilon}(\zeta)\,u\right) \otimes v_{\epsilon}.
\end{eqnarray*}
Then 
\begin{eqnarray}
&& 
{}_{(i)}\bra{{\rm vac}}
(E_{-\epsilon_{1}, \overline{\epsilon}_{1}})_{1} \cdots 
(E_{-\epsilon_{n}, \overline{\epsilon}_{n}})_{n}
\ket{{\rm vac}}_{(i)}
\label{eq:correlation-trace} \\ 
&& {}=\chi^{-1}g^{n} \,  
{\rm tr}_{V(\Lambda_{i})}\left( q^{2D^{(i)}}
\Phi_{\epsilon_{n}}(q^{-1}\zeta_{n}) \cdots
\Phi_{\epsilon_{1}}(q^{-1}\zeta_{1})  
\Phi_{\overline{\epsilon}_{1}}(\zeta_{1}) \cdots 
\Phi_{\overline{\epsilon}_{n}}(\zeta_{n}) 
\right), 
\nonumber
\end{eqnarray}
where $\zeta_{j}=q^{\lambda_j}$, 
$D^{(i)}=-(\Lambda_{0}+\Lambda_{1})+\frac{i}{2}$, and 
\begin{eqnarray*}
\chi=\frac{1}{(q^{2}; q^{4})_{\infty}}, \quad 
g=\frac{(q^{2}; q^{4})_{\infty}}{(q^{4}; q^{4})_{\infty}}. 
\end{eqnarray*}

The vertex operators satisfy the following relations: 
\begin{eqnarray}
&& 
\Phi_{\epsilon_{2}}(\zeta_{2}) \Phi_{\epsilon_{1}}(\zeta_{1})=
\sum_{\epsilon_{1}', \epsilon_{2}'=\pm} 
R_{JM}(\zeta_{1}/\zeta_{2})_
{\epsilon_{1}\epsilon_{2}}^{\epsilon_{1}'\epsilon_{2}'}
\Phi_{\epsilon_{1}'}(\zeta_{1}) \Phi_{\epsilon_{2}'}(\zeta_{2}), 
\label{eq:VOrel1} \\ 
&& 
\xi^{D^{(1-i)}} \cdot \Phi_{\epsilon}(\zeta) \cdot 
\xi^{-D^{(i)}}=\Phi_{\epsilon}(\xi\zeta), \\
&& 
g\sum_{\epsilon=\pm}\Phi_{-\epsilon}(\zeta)\Phi_{\epsilon}(q\zeta)=
{\rm id}.
\label{eq:VOrel2}
\end{eqnarray}
The basic relations 
\eqref{eq:rqkz1}--\eqref{eq:rqkz2} 
are simple consequences of 
\eqref{eq:correlation-trace} and \eqref{eq:VOrel1}--\eqref{eq:VOrel2}.

\section{The scalar factors}\label{app:rho}

We collect here formulas for the scalar factors
which enter the definition of the $L$-operators. 
\bigskip

\noindent\underline{$XXZ$ case}
\bigskip

$\bullet$ {Massless regime} ($0<\nu<1$)

\bea
\rho(\la,d)=-
\frac{S_2\bigl(1-\frac{d}{2}-\lambda\bigr)}
{S_2\bigl(1-\frac{d}{2}+\lambda\bigr)}
\frac{S_2\bigl(2-\frac{d}{2}+\lambda)}
{S_2\bigl(2-\frac{d}{2}-\lambda)},
\label{eq:rho_trig1}
\ena
where $S_2(\la)=S_2(\la|2,1/\nu)$ stands for the
double sine function. 
\medskip

$\bullet$ {Massive regime} ($\nu\in i\R_{>0}$)

\bea
\rho(\la,d)=-\z
\frac{(q^{2-d}\z^{-2})_\infty}{(q^{2-d}\z^2)_\infty}
\frac{(q^{4-d}\z^{2})_\infty}{(q^{4-d}\z^{-2})_\infty}
\label{eq:rho_trig2}
\ena
where $(z)_\infty=\prod_{j=0}^\infty(1-zq^{4j})$. 

\bigskip

\noindent\underline{\bf $XYZ$ case}
\bigskip

$\bullet$ {Disordered regime} ($\eta,t\in i\R$, $-i\eta>0$)
\bea
&&\rho(t,d)=-e^{2\pi i t}
\frac{\gamma((4-d)\eta-2t)}{\gamma((4-d)\eta+2t)}
\frac{\gamma((2-d)\eta+2t)}{\gamma((2-d)\eta-2t)},
\label{eq:rho_ell}
\ena
where $\gamma(u)=\Gamma(u,4\eta,\tau)$ and 
\be
\Gamma(u,\sigma, \tau):=
\prod_{j,k=0}^\infty
\frac{1-e^{2\pi i((j+1)\tau+(k+1)\sigma-u)}}
{1-e^{2\pi i(j\tau+k\sigma+u)}}
\en
denotes the elliptic Gamma function. 

\medskip

$\bullet$ {Ordered regime} ($\eta,t\in \R$, $\eta<0$)
\be
&&\rho(t,d)=
e^{-\frac{4\pi i}{\tau}(d-1)\eta t}
\times \rho'(t',d),
\en
where $\rho'(t',d)$ is obtained from \eqref{eq:rho_ell} 
by replacing $t,\eta,\tau$ with 
\be
t'=\frac{t}{\tau},\quad 
\eta'=\frac{\eta}{\tau},
\quad \tau'=-\frac{1}{\tau},
\en
respectively. 

\bigskip

\bigskip

\noindent
{\it Acknowledgments.}\quad
Research of HB is supported 
by the RFFI grant \#04-01-00352.
Research of MJ is 
supported by 
the Grant-in-Aid for Scientific Research B2--16340033
and 
A2--14204012.
Research of TM is
supported by 
the Grant-in-Aid for Scientific Research A1--13304010.
Research of FS is 
supported by INTAS grant \#03-51-3350, %and by %\#00-00055 
EC networks  "EUCLID", contract number HPRN-CT-2002-00325, 
"ENIGMA", contract number MRTN-CT-2004-5652, and 
GIMP program (ANR), contract number ANR-05-BLAN-0029-01.
Research of YT is supported by Grant-in-Aid for 
Young Scientists (B) No.\,17740089. 
This work was also supported by the grant of 21st Century 
COE Program at Graduate School of Mathematical Sciences, 
the University of Tokyo,
and at RIMS, Kyoto University. 

%MJ and HB would like to thank 
%F. G{\"o}hmann, 
%C. Korff, 
%A. Kl{\"u}mper,  
%J.M. Maillet, 
%J. Suzuki 
%and 
%V. Terras 
%for interests and discussions. 
%HB is grateful to P. Pyatov for explanation about $q$-traces. 
%
%MJ and HB would like to thank 
%F. G{\"o}hmann, A. Kl{\"u}mper and J. Suzuki 
%for interests and discussions. 
%MJ also thanks 
%C. Korff, J.M. Maillet and V. Terras for discussions. 
%HB is grateful to P. Pyatov for explanation about $q$-traces. 

MJ and HB would like to thank F. G{\"o}hmann and A. Kl{\"u}mper 
for interests and discussions. 
MJ also thanks C. Korff, J.M. Maillet and V. Terras for discussions. 
HB is grateful to J. Suzuki for discussions and to P. Pyatov for 
explanations about $q$-traces.


\bigskip


\begin{thebibliography}{[FJKLM]}

\bibitem{Bethe} 
H.~Bethe,
\newblock Zur Theorie der Metalle. I.
Eigenwerte und Eigenfunktionen der linearen Atomkette,
\newblock{Zeitschrift f\"ur Physik} {\bf 71} (1931) 205.

\bibitem{BJMST1} 
H.~Boos, M.~Jimbo, T.~Miwa, F.~Smirnov and Y.~Takeyama,  
\newblock 
A recursion formula for the correlation 
functions of an inhomogeneous XXX model,
\newblock {\em Algebra and Analysis} {\bf 17}
(2005), 115--159.

\bibitem{BJMST2} 
H.~Boos, M.~Jimbo, T.~Miwa, F.~Smirnov and
Y.~Takeyama,  
\newblock Reduced qKZ equation and correlation
functions of the XXZ model,
\newblock {\em Commun. Math. Phys.} {\bf 261} (2006), 245-- 276.

\bibitem{BJMST3} 
H.~Boos, M.~Jimbo, T.~Miwa, F.~Smirnov and Y.~Takeyama,  
\newblock Traces on the Sklyanin algebra and 
correlation functions of the eight-vertex model,
\newblock {\em J. Phys. A: Math. Gen.} {\bf 38} (2005), 7629-7659.

\bibitem{BJMST4} 
H.~Boos, M.~Jimbo, T.~Miwa, F.~Smirnov and Y.~Takeyama,  
\newblock 
Density matrix of a finite sub-chain of 
the Heisenberg anti-ferromagnet, 
\newblock hep-th/0506171, to appear in Lett. Math. Phys. 

\bibitem{BK1} 
H.~Boos and V.~Korepin, 
\newblock Quantum spin chains and Riemann zeta functions with odd arguments, 
\newblock hep-th/0104008,
{\em J. Phys. }{\bf A 34} (2001), 5311--5316.

\bibitem{BK2} 
H.~Boos and V.~Korepin, 
\newblock Evaluation of integrals representing correlations
in XXX Heisenberg spin chain, 
\newblock in \emph{MathPhys Odessey 2001}, Birkh{\"a}user
(2001), 65--108. 

\bibitem{BKS} 
H.~Boos, V.~Korepin and F.~Smirnov, 
\newblock Emptiness formation probability and 
quantum Knizhnik-Zamlodchikov equation, 
\newblock hep-th/0209246, 
{\it Nucl. Phys. B} Vol. 658/3 (2003), 417 --439.

\bibitem{BKS2} 
H.~Boos, V.~Korepin and F.~Smirnov, 
\newblock Connecting lattice and relativistic models via
conformal field theory, 
\newblock math-ph/0311020.

\bibitem{GKS}
F.~G{\"o}hmann, A.~Kl{\"u}mper and A.~Seel, 
\newblock 
Integral representations for correlation functions of the 
XXZ chain at finite temperature,  
\newblock 
{\em J. Phys. A} {\bf 38} (2005), 1833--1842.

\bibitem{JMMN} M.~Jimbo, T.~Miwa, K.~Miki
and A.~Nakayashiki, 
\newblock Correlation functions of the XXZ model
for $\Delta<-1$,
\newblock {\em Phys. Lett. A} {\bf 168} (1992), 256--263.

\bibitem{JM}
M.~Jimbo and T.~Miwa, 
\newblock
Algebraic analysis of solvable lattice models, 
\newblock {\it Reg. Conf. Ser. in Math. \,}{\bf 85 }, {1995}. 

\bibitem{JM2} 
M.~Jimbo and T.~Miwa, 
\newblock 
Quantum Knizhnik-Zamolodchikov equation at $|q|=1$ and 
correlation functions of the XXZ model in the gapless regime,
\newblock 
{\em J. Phys.} \textbf{A 29} (1996), 2923--2958. 

\bibitem{KMT}
N.~Kitanine, J.-M.~Maillet and V.~Terras,
\newblock 
Correlation functions of the XXZ Heisenberg 
spin-$\frac{1}{2}$-chain in a magnetic field,
\newblock 
{\em Nucl. Phys. B} \textbf{567} (2000), 554--582. 

\bibitem{KMST} N.~Kitanine, J.~M.~Maillet, N.~A.~Slavnov and V.~Terras, 
\newblock Dynamical correlation functions of the XXZ spin-1/2 chain,
\newblock hep-th/0407223. 

\bibitem{KMST1}
N.~Kitanine, J.-M.~Maillet, N.~Slavnov and V.~Terras,
\newblock 
On the algebraic Bethe Ansatz approach to the correlation functions 
of the XXZ spin-1/2 Heisenberg chain, 
\newblock hep-th/0505006.

\bibitem{KMST2}
N.~Kitanine, J.-M.~Maillet, N.~Slavnov and V.~Terras,
\newblock 
Large distance asymptotic behavior of the emptiness
formation probability of the XXZ spin-$1/2$ Heisenberg chain, 
\newblock 
{\em J. Phys. A} \textbf{35} (2002),  L753.

\bibitem{KRS} P.P.~Kulish, N.Yu.~Reshetikhin and E.K.~Sklyanin, 
\newblock Yang-Baxter equation and representation theory. I, 
{\em Lett. Math. Phys.} {\bf 5} (1981), 393--403. 

\bibitem{KLNS}
V.~Korepin, S.~Lukyanov, Y.~Nishiyama and M.~Shiroishi, 
\newblock 
Asymptotic behavior of the emptiness formation probability, 
\newblock 
{\em Phys. Lett. A} {\bf 312} (2003), 21--26. 

\bibitem{LS} A. LeClair, F. Smirnov,
\newblock 
Infinite quantum group symmetry of fields in massive $2$D quantum field theory, 
\newblock {\it Int. J. Mod. Phys. A} {\bf 7} (1992), 2997--3022.

\bibitem{PM} P.~Martin, 
\newblock 
\emph{Potts models and related problems in 
statistical mechanics}, 
\newblock World Scientific, Singapore, 
1991.

\bibitem{PS}V. Pasquier, H. Saleur, 
\newblock 
Common structures between finite systems and 
conformal field theories through quantum groups,
\newblock 
{\it Nucl. Phys. B} {\bf  330} (1990), 523--556.

\bibitem{RS}
N.~Reshetikhin and F.~Smirnov,
\newblock Hidden quantum group symmetry and integrable 
perturbations of conformal field theory,
\newblock {\em Commun. Math. Phys.} {\bf 131} (1990), 
157--177.

\bibitem{KSTS} G.~Kato, M.~Shiroishi, M.~Takahashi, K.~Sakai, 
\newblock
Next Nearest-Neighbor Correlation Functions of the Spin-1/2 XXZ
Chain at Critical Region,
\newblock {\it J. Phys. A: Math. Gen.} {\bf 36} (2003) L337. 

\bibitem{SSNT} K.~Sakai, M.~Shiroishi, Y.~Nishiyama and M.~Takahashi, 
\newblock
Third neighbor correlators of spin-1/2 Heisenberg antiferromagnet, 
\newblock {\it Phys. Rev. E} {\bf 67} (2003), 065--101.

\bibitem{KSTS2} G.~Kato, M.~Shiroishi, M.~Takahashi, K.~Sakai,
\newblock
Third-neighbor and other four-point correlation functions of spin-1/2
XXZ chain,
\newblock {\it J. Phys. A: Math. Gen.} {\bf 37} (2004) 5097. 

\bibitem{SST}
J.~Sato, M.~Shiroishi and M.~Takahashi, 
\newblock 
Correlation functions of the spin-$1/2$ 
anti-ferromagnetic Heisenberg chain:
exact calculation via the generating function,  
\newblock {\it Nucl, Phys. B} {\bf 729} (2005), 441--466.

\bibitem{Sk} E.~K.~Sklyanin, 
\newblock Some algebraic structures connected with the Yang-Baxter
equation, 
\newblock {\em Func. Anal. and Appl. }{\bf 16} (1982), 27--34;
{\bf 17} (1983), 34--48.

\bibitem{Tak} M.~Takahashi,
\newblock 
Half-filled Hubbard model at low temperature,
\newblock 
{\em J. Phys. C} \textbf{10} (1977) 1298. 


\end{thebibliography}
\end{document}